\pdfoutput=1
\documentclass[nonacm,sigplan]{acmart}
\acmSubmissionID{388}

\usepackage{def} %% put new commands here

\usepackage{hyperref}

\hypersetup{
  colorlinks,
  linkcolor={blue},
  citecolor={red!70!black},
  urlcolor={blue!70!black}
}

\usepackage[compact]{titlesec}
\usepackage{printlen}
\usepackage{pifont}
\usepackage{multirow}
\usepackage{wrapfig}
\usepackage{url}
\usepackage{xcolor}
\usepackage{enumitem}
\usepackage{capt-of}
\usepackage{booktabs}
\usepackage{subcaption}
\usepackage{calrsfs}
\usepackage{listings}

 \usepackage[htt]{hyphenat}

\usepackage{xspace}

\usepackage[noabbrev]{cleveref}

\newcommand{\eg}{{\it e.g.}, }
\newcommand{\ie}{{\it i.e.}, }

\newcommand{\blit}{\texttt{BloxDataLoader}\xspace}
\newcommand{\blcl}{\texttt{BloxClientLibrary}\xspace}
\newcommand{\wmanager}{\texttt{WorkerManager}\xspace}
\newcommand{\cscheduler}{\texttt{CentralScheduler}\xspace}

\newcommand{\xmark}{\ding{55}}%

\newcommand{\phtrace}{\texttt{Philly-Trace}\xspace}
\newcommand{\potrace}{\texttt{Pollux-Trace}\xspace}
\newcommand{\titrace}{\texttt{Tiresias-Trace}\xspace}

\newcommand\blfootnote[1]{%
  \begingroup
  \renewcommand\thefootnote{}\footnote{#1}%
  \addtocounter{footnote}{-1}%
  \endgroup
}
\usepackage{graphicx}
\graphicspath{ {./images/} }

\newenvironment{myitemizeleft}
{
   \vspace{0pt}
    \begin{list}{$\bullet$ }{\leftmargin=1em \itemindent=0em}
        \setlength{\topsep}{0em}
        \setlength{\parskip}{0pt}
        \setlength{\partopsep}{0pt}
        \setlength{\parsep}{0pt}
        \setlength{\itemsep}{0.1mm}
}
{
    \end{list}
}

\definecolor{brewerpurple}{HTML}{AF4EA3}
\definecolor{brewerblue}{HTML}{377EB8}
\definecolor{NavyBlue}{HTML}{006EB8}
\definecolor{BrickRed}{HTML}{B6321C}
\definecolor{ForestGreen}{HTML}{009B55}

\acmYear{2024}\copyrightyear{2024}
\acmConference[EuroSys '24]{European Conference on Computer Systems}{April 22--25, 2024}{Athens, Greece}
\acmBooktitle{European Conference on Computer Systems (EuroSys '24), April 22--25, 2024, Athens, Greece}
\acmPrice{15.00}
\acmDOI{10.1145/3627703.3629583}
\acmISBN{979-8-4007-0437-6/24/04}

\lstdefinestyle{customc2}{
    emph={update\_metrics, prune\_completed\_jobs, exec\_jobs, accept, schedule, place, AcceptAll, Fifo, Consolidated, pop\_wait\_queue, update\_cluster},
    emphstyle=\bfseries\color{NavyBlue},
    commentstyle=\color{ForestGreen}\itshape\ttfamily,
       morekeywords={yield},
    stringstyle=\color{red}\ttfamily,
    emph={[2]admission\_policy, scheduling\_policy, placement\_policy},emphstyle={[2]\bfseries\color{BrickRed}},
    language=Python,
    keywordstyle=\bfseries\color{green!40!black}
}

\begin{document}
\title{\sysname{}: A Modular Toolkit for Deep Learning Schedulers }
\author{Saurabh Agarwal}
\affiliation{University of Wisconsin-Madison
\country{}}
\authornote{Microsoft Research Intern}
\author{Amar Phanishayee}
\affiliation{Microsoft Research
\country{}}
\author{Shivaram Venkataraman}
\affiliation{University of Wisconsin-Madison
\country{}}

\begin{abstract}
Deep Learning (DL) workloads have rapidly increased in popularity in enterprise clusters and several new cluster schedulers have been proposed in recent years to support these workloads.
With rapidly evolving DL workloads, it is challenging to quickly prototype and compare scheduling policies across workloads. Further, as prior systems target different aspects of scheduling (resource allocation, placement, elasticity etc.), it is also challenging to combine these techniques and understand the overall benefits.

To address these challenges we propose \system, a modular toolkit which allows developers to compose individual components and realize diverse scheduling frameworks.
We identify a set of core abstractions for DL scheduling, implement several existing schedulers using these abstractions, and verify the fidelity of these implementations by reproducing results from prior research.
We also highlight how we can evaluate and compare existing schedulers in new settings: different workload traces, higher cluster load, change in DNN workloads and deployment characteristics.
Finally, we showcase \sysname{}'s extensibility by composing policies from different schedulers, and implementing novel policies with minimal code changes. \sysname{} is available at \url{https://github.com/msr-fiddle/blox}.
\end{abstract}
\maketitle
\pagestyle{plain}

\section{Introduction}

Modern deep neural networks (DNNs) are increasingly used in enterprises to solve a range of problems such as image classification~\cite{krizhevsky2012imagenet,he2016resnet}, semantic segmentation~\cite{sun2019HRNetV2}, image generation~\cite{goodfellow2014generative}, translation~\cite{wu2016google, raffel2019t5}, and language modeling~\cite{vaswani2017attention,devlin2018bert,shoeybi2019megatron,radford2019gpt2,brown2020language}.
These workloads pose new demands when compared to big-data workloads, such as in MapReduce~\cite{dean2008mapreduce} or Spark~\cite{zaharia2010spark}, along a number of dimensions.
DNN jobs are not made up of short diverse tasks but instead are long-running jobs with repeated iterations over different input data items. Thus, DNN jobs have different granularities for job preemption, have sophisticated application-specific metrics for termination (training loss) and elasticity (training progress), and have multi-dimensional resource requests both along newer dimensions of compute acceleration (e.g., TPUs or GPUs) as well as traditional resource types (compute, memory, interconnects).
Given the prevalence and importance of these workloads there has been a large body of recent research that has proposed schedulers to support and exploit the unique characteristics of these jobs~\cite{qiao2021pollux, gu2019tiresias, xiao2018gandiva, mohan2021synergy, mahajan2020themis, peng2018optimus, narayanan2020heterogeneity, jeon2019analysis, hwang2021elastic, crankshaw2017clipper, shen2019nexus, gujarati2020serving, hwang2021elastic, xiao2020antman, le2020allox, weng2022mlaas}.

Analyzing trends across deep learning (DL) schedulers, we observe that

while each prior work proposes new innovations for DL scheduling, their contributions are typically focused on a narrow part of the scheduler stack e.g., new resource allocation policies~\cite{peng2018optimus, qiao2021pollux, xiao2020antman, mohan2021synergy}, handling elasticity~\cite{xiao2020antman,qiao2021pollux}, or placement policies~\cite{qiao2021pollux,peng2018optimus, gu2019tiresias}. However, authors have to either develop an entirely new scheduler stack (e.g., Gavel~\cite{narayanan2020heterogeneity}) or target their policies to a specific enterprise stack (e.g., HiveD~\cite{zhao2020hived} in PAI~\cite{mspai} from Microsoft, Pollux~\cite{qiao2021pollux} in AdaptDL \cite{adaptdl} from Petuum, etc.).

Having each scheduler use a different stack makes it \textbf{\emph{challenging to compare, compose, or re-evaluate innovations}}.
The increase in the popularity of DNNs, and consequently cluster load, necessitates comparing existing schedulers to answer questions such as: \emph{how do previously proposed scheduling policies compare to each other on newer cluster traces or higher cluster loads, evaluated on a common footing?}

Re-evaluation of scheduling policies is also necessitated by workload evolution. The rapid evolution of DNN workloads has seen popular DNN architectures evolve from CNNs to RNNs to Transformer-based models~\cite{jouppi2020domain}. Thus, it becomes necessary to re-evaluate scheduler efficacy; for example, to answer questions such as: \emph{how effective is the placement policy proposed in Tiresias~\cite{gu2019tiresias} for newer models or deployments?}
Further, it is also challenging to \emph{compose} contributions of different schedulers to evaluate their overall impact.
For example, \emph{how effective is composing aggressive admission control with a scheduling policy that aims for fairness across jobs?}

We also observe that DL scheduling policies are designed to benefit particular arrival patterns. However, often these patterns do not hold over long periods of time, \eg there might be a lot of short jobs during working hours when ML engineers are testing their code, but nights and weekends are dominated by long running jobs. This indicates that users might benefit if the scheduling policies evolve based on arrival patterns, job types etc. However, designing such dynamic policy changes is challenging in current scheduler architectures.

\blfootnote{To appear in the European Conference on Computer Systems (EuroSys '24)}
\noindent\textbf{Contributions.}  In this paper we propose a toolkit that can help answer the above questions. We present, \sysname{}, a new scheduler toolkit with a set of clean, modular abstractions and implementations.  \system{} can be used to compare and understand existing DL schedulers (re-visiting the past in new light), and our abstractions also serve as building blocks for researchers to realize new
scheduler designs (looking into the future).
In this pursuit, we are directly inspired by two iconic systems research toolkits from the past: the FluxOS toolkit~\cite{ford-hotos97-flux, ford-sosp97-flux} for operating systems research and the Click toolkit ~\cite{Kohler-TOCS2000-Click, Kohler-SOSP99-Click} for flexible and configurable routers.

\begin{figure*}[t]
    \includegraphics[width=0.9\linewidth]{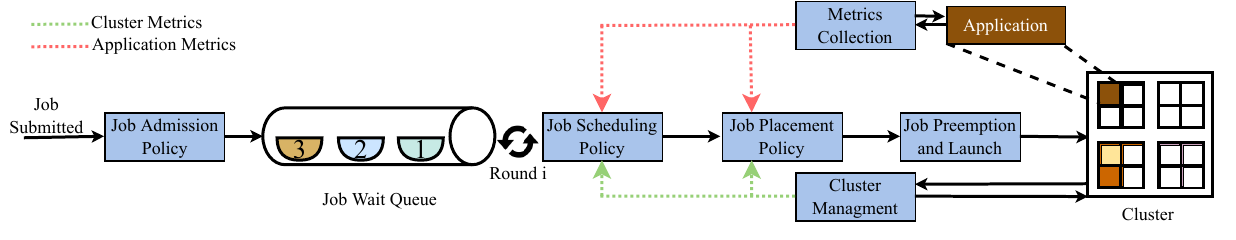}
    \centering
    \vspace{-15pt}
    \caption{\small{\textbf{DL Scheduler workflow in \system}: Key abstractions, and their high-level interactions, required for building DL schedulers.}}
    \label{fig:sched_schem}
    \vspace{-15pt}
\end{figure*}

By analyzing prior schedulers we identify \emph{seven key abstractions} that can be \emph{composed} to realize a diverse set of DL schedulers.
Figure~\ref{fig:sched_schem} shows a schematic overview of a generic DL scheduler in \system{} highlighting these abstractions and their interactions.

We implement concrete instances of these abstractions and compose them to realize seven existing cluster schedulers including FIFO, Tiresias~\cite{gu2019tiresias}, Optimus~\cite{peng2018optimus}, Themis~\cite{mahajan2020themis}, Gavel~\cite{narayanan2020heterogeneity}, Pollux~\cite{qiao2021pollux}, and Synergy~\cite{mohan2021synergy}.
Additionally, we also validate that our implementation of prior schedulers are accurate by reproducing some of their experiments; we compare the results from the \sysname{} implementation of these schedulers with their reported numbers or results from running their open source implementations.

Using our toolkit we also conduct a number of case studies that showcase how \system{} can be used to glean new insights about DL scheduling.
By varying cluster load, we show the differences in how existing scheduling policies~\cite{qiao2021pollux,peng2018optimus, gu2019tiresias} handle the trade-off between average job completion time (JCT) and responsiveness (\S\ref{case-study:comparison}) at high load (at which many of these schedulers were not evaluated before).
We also study how changes to workloads and cluster hardware necessitate re-evaluating placement policies and our findings show the importance of using accurate profiles for placement (\S\ref{case-study:consolidation}).

Furthermore, to showcase the extensibility of \sysname{}, we also investigate the ease of developing new scheduling policies and scheduler mechanisms for DL training.
First, to demonstrate the ease of composing modules in \system{}, we show how combining aggressive admission control with a fair-scheduling policy (LAS) can help alleviate the problem of slow job progress (unreasonably large JCTs) caused due to frequent preemptions at high load (\S\ref{case-study:composition}).
Next, we extend our composition based approach to \emph{automatically synthesize} DL schedulers based on the observed workload. This novel Automatic Scheduler Synthesizer, identifies the set of polices which provide maximum improvement for a user selected metric and uses simulation to automatically switch between policies.
Finally, we also develop a loss-based job termination feature that can proactively free up resources when model training has converged.

We also validate the usability and reproducibility of simulations in \sysname{}. The modular design of \sysname{} ensures that only two modules need to be modified between simulations and cluster runs, and we verify that \sysname{} simulations match real executions (JCT within 6.1\% on average) using a %32
GPU cluster on AWS. To validate the usability of \sysname{} we also discuss the results from a study where two groups of students reproduced results from Themis~\cite{mahajan2020themis} and Optimus~\cite{peng2018optimus} as a part of their class projects. %  \amar{fix this last sentence.}
We hope to make \sysname{} a resource that the systems research community can use to accelerate the development of new scheduler research targeting DL jobs and have released \sysname{} as an open source project at \url{https://github.com/msr-fiddle/blox}.

\begin{comment}
\textbf{Roadmap.}
%In the rest of this paper,
We first present motivation and background on DL schedulers (\S\ref{sec:motivation}) and then provide a high-level overview of \sysname{} including its key abstractions and how to compose them (\S\ref{sec:design}).
We validate \sysname{}'s efficacy in implementing new and existing schedulers; we present case studies for implementing known policies, composing existing ones, evaluating existing schedulers on newer workloads and deployments, and adding novel policies (\S\ref{sec:case-studies}).
We then present low-level design and implementation details and evaluate the fidelity of \sysname{}'s simulator (\S\ref{sec:impl}).
While scheduling of DL training jobs is our primary goal,  we also discuss how our abstractions are general enough to support schedulers for inference~\cite{shen2019nexus} and hyper parameter tuning jobs~\cite{li2017hyperband} (\S\ref{sec:discussion}), before concluding (\S\ref{sec:conclusion}).
\end{comment}

% \the\linewidth
\section{Background and Motivation}
\label{sec:motivation}
We motivate the unique challenges in scheduling DL training jobs and provide an overview of existing DL schedulers.

\subsection{Cluster Schedulers}
A rich line of research developed scheduling frameworks like SLURM~\cite{yoo2003slurm}, YARN~\cite{vavilapalli2013yarn}, Mesos~\cite{hindman2011mesos}, Kubernetes~\cite{kubernetes} and Borg~\cite{verma2015large} which are widely used for scheduling high performance computing jobs, big-data jobs or long running internet services like HTTP servers. However, they are not sufficient for DL training jobs because of two main reasons.
First, schedulers like Mesos and YARN handle large big-data jobs that are composed of several short-running tasks or long running internet services that run at high priority and thus are usually never preempted. On the other hand, DL jobs are usually long running with their computation being repeated for a large number of iterations. Therefore, DL schedulers unlike big data schedulers need to frequently preempt a running job to prevent ``head-of-line-blocking'' for better resource management~\cite{xiao2018gandiva}.
Second, DL schedulers often need access to application level metrics like loss, gradient norm, throughput, etc., to support DL-specific aspects like finish-time fairness~\cite{mahajan2020themis} or gradient-based elasticity~\cite{qiao2021pollux}, which is not easily available in existing scheduling frameworks. Thus, while prior DL schedulers~\cite{zhao2020hived,xiao2020antman, qiao2021pollux} have been implemented as plugins on Kubernetes~\cite{kubernetes} or YARN~\cite{vavilapalli2013yarn}, these systems typically need to design additional DL-specific features to support iteration-level preemption or app-level metric collection.

Developing and deploying DL schedulers is also complicated by rapid evolution of DL workloads, \eg while CNN models like VGG16 and ResNet50 were widely used a few years ago, industry reports~\cite{jouppi2020domain} show that Transformer-based models such as BERT and deep learning based recommendation models (DLRM)~\cite{acun2021understanding} now form a significant portion of the enterprise ML workload. Further, newer hardware such as TPUs (or newer generation of GPUs) also necessitate new mechanisms for scheduling.
This rapid evolution of workload and hardware motivates the need for scheduling frameworks to support quick prototyping of new policies.

Several prior works have studied schedulers, metrics to evaluate schedulers and different workloads. Verma et. al studied metrics for evaluating schedulers on data-centers workloads~\cite{verma2014evaluating}.
Amvrosiadis et. al consider traces from large HPC clusters to highlight the importance of dataset plurality in job scheduling research~\cite{amvrosiadis2017bigger}

In \S\ref{sec:case-studies} we also study different metrics and performance of schedulers on different types of traces, albeit our focus is solely on DL schedulers.

\subsection{Deep Learning Schedulers}
\label{sec:comps_required}

Unlike the task-based scheduling approach used by schedulers such as Mesos and YARN, DL schedulers are \emph{round based}, \ie after a fixed interval (round length) they make scheduling decisions regarding the jobs to run often requiring preempting in-progress jobs, thus neccessating the need for checkpointing and preemption of jobs and resuming from the checkpoints.
Round based scheduling has been shown to be necessary for achieving good cluster efficiency, low queuing times and avoiding head-of-line blocking~\cite{xiao2018gandiva,gu2019tiresias,mahajan2020themis,narayanan2020heterogeneity}.

Most prior work in DL scheduling is focused on developing policies that can improve a number of metrics including
 job completion time (JCT)~\cite{qiao2021pollux, gu2019tiresias, xiao2018gandiva, mohan2021synergy}, makespan~\cite{gu2019tiresias, xiao2018gandiva}, cluster utilization~\cite{qiao2021pollux, peng2018optimus}, throughput~\cite{qiao2021pollux, gu2019tiresias, xiao2018gandiva, mohan2021synergy} and fairness~\cite{mahajan2020themis,gandivafair}.
These scheduling policies are typically invoked at the end of every round to decide which jobs should be selected to run in the next round and how many resources should be allocated to each selected job. Since DL training jobs are also known to be placement sensitive~\cite{gu2019tiresias}, some schedulers also use additional placement policies to decide which machine in the cluster will run this job.

To perform scheduling, DL schedulers use a number of system-level and application level metrics. Schedulers such as Gavel~\cite{narayanan2020heterogeneity}, Gandiva ~\cite{xiao2018gandiva}, and Synergy~\cite{mohan2021synergy} use system level metrics like GPU memory usage, DRAM usage, etc., to take scheduling decisions.
A number of other schedulers also use application level metrics like per iteration time~\cite{narayanan2020heterogeneity, gu2019tiresias, peng2018optimus} or training progress~\cite{qiao2021pollux, peng2018optimus}.

We observe that the structure and the high level components are broadly similar across DL schedulers. It is only the internals of the components that change, \eg all existing schedulers need some metrics like GPU usage, throughput, gradient noise, etc., to make scheduling decisions and the only change across schedulers is in what metrics are required. This insight helps us develop a set of abstractions required for DL scheduling which we describe in \S~\ref{sec:blox_overview} .

\subsection{Need for a modular framework}
The current scheduler landscape consists of a plethora of research schedulers with each having their own specific software stack. This makes it challenging to compare, compose, or re-evaluate innovations across schedulers, and eventually affects adoption of new techniques, as cluster operators are unable to convince themselves of the efficacy of individual innovations on a common footing.  We believe this lack of interoperability stems from a lack of clear specifications for various scheduler modules, their interfaces, and modes of interaction.  Based on our experience building research schedulers over the years, studying large-scale deployments, and speaking to users and operators of production clusters, in the next couple of sections we highlight a simple set of clearly defined abstractions for DL schedulers. We show how these abstractions can enable reproducibility, easy interoperability and comparison, re-evaluating contributions of existing schedulers on newer hardware or workload traces, and easy addition of novel scheduling ideas.

\section{Blox Overview}
\label{sec:blox_overview}
\begin{figure}
% \begin{minipage}[b]{0.5\textwidth}
\centering
\resizebox{0.95\linewidth}{!}{
\begin{lstlisting}[style=customc2,escapechar=|]
 from blox import ClusterState, JobState, BloxManager

 def main(args):
     admission_policy = admission_control.AcceptAll(args)|\label{line:ap}|
     scheduling_policy = schedulers.Fifo(args)|\label{line:sp}|
     placement_policy = placement.Consolidated(args)|\label{line:pp}|
    
     blox_mgr = BloxManager(args)
     cluster_state = ClusterState(blox_mgr, args)|\label{line:cs}|
     job_state = JobState(blox_mgr, args)|\label{line:js}|

     while not blox_instance.terminate:|\label{line:sc-start}|
         # update set of active machines
         blox_mgr.update_cluster(cluster_state)
         # update metrics of all jobs run in the 
         # previous round (including failed jobs)
         blox_mgr.update_metrics(cluster_state, job_state)
         blox_mgr.prune_completed_jobs(
             cluster_state, job_state)

         # retrieve new jobs from wait queue
         new_jobs = blox_mgr.pop_wait_queue(args.simulate)|\label{line:pop}|
         # acceptance policy
         accepted_jobs = admission_policy.accept(new_jobs, 
             cluster_state, job_state)|\label{line:accept}|
         # update runnable jobs with accepted jobs
         job_state.add_new_jobs(accepted_jobs)
        
         # get new job schedule
         new_job_schedule = scheduling_policy.schedule(
             job_state, cluster_state)|\label{line:schedule}|
         # where to launch
         to_launch, to_suspend = placement_policy.place(
             new_job_schedule, cluster_state, job_state)|\label{line:place}|
         # launch jobs
         blox_mgr.exec_jobs(
             to_launch, to_suspend, cluster_state, job_state)|\label{line:exec}|
         # wait until next round
         if not args.simulate: |\label{line:sleep}|
             time.sleep(args.round_duration)|\label{line:sc-end}|
\end{lstlisting}}
%\lstinputlisting[language=pseudocode]{codes/blox_new_flow.tex}}
\caption{\small{\textbf{Blox Flow:} The code above shows a simplified example of how to chain abstraction to easily build a scheduler in \system{}.}}
\label{fig:blox_chain_flow}
\vspace{-0.2in}
\end{figure}

\begin{table*}[t]
\caption{\small{\textbf{Abstractions and their instances as used by DL schedulers:} We observe that following abstractions can be used to build a large range of DL schedulers. An interesting observations is that there is a significant amount of overlap in the instances of abstractions used across several DL schedulers. }}
\vspace{-10pt}
\label{tab:abstraction_used}
\resizebox{\linewidth}{!}{
\begin{tabular}{@{}lcccccc@{}}
\toprule
\textbf{Abstraction}                                      & \textbf{Tiresias}                                    & \textbf{Optimus}                                       & \textbf{Themis}                                                                                        & \textbf{Gavel}                                        & \textbf{Pollux}                                        & \textbf{Synergy}                                                                                           \\ \midrule
\multicolumn{1}{l|}{Job Admission Policy}       & \multicolumn{6}{c}{FIFO admission}                                                                                                                                                                                                                                                                                                                                                          \\ \midrule
\multicolumn{1}{l|}{Cluster Management}         & \multicolumn{6}{c}{Add new nodes, Collect cluster metrics (CPU/GPU compute usage, CPU/GPU memory usage, Disk usage), Detect failures, Removed failed nodes}                                                                                                                                                                                                                                                                                                                              \\ \midrule
\multicolumn{1}{l|}{Job Scheduling Policy}      & \multicolumn{1}{c|}{discreet LAS}           & \multicolumn{1}{c|}{largest marginal gain}    & \multicolumn{1}{c|}{finish time fair policy}                                                  & \multicolumn{1}{c|}{heterogeneity aware LAS} & \multicolumn{1}{c|}{max mean speedup}         & \multicolumn{1}{c}{resource sensitive FIFO}                                                          \\ \midrule
\multicolumn{1}{l|}{Job Placement Policy}       & \multicolumn{1}{c|}{application determined} & \multicolumn{1}{c|}{min communication}        & \multicolumn{1}{c|}{application determined}                                                   & \multicolumn{1}{c|}{maximize consolidation}  & \multicolumn{1}{c|}{min network interferance} & \multicolumn{1}{c}{greedy resource allocation}                                                                      \\ \midrule
\multicolumn{1}{l|}{Job Launch Mechanism}       & \multicolumn{6}{c}{command line}                                                                                                                                                                                                                                                                                                                                                            \\ \midrule
\multicolumn{1}{l|}{Job Preemption and restart} & \multicolumn{6}{c}{Iteration-boundary checkpoint based}                                                                                                                                                                                                                                                                                                                           \\ \midrule
\multicolumn{1}{l|}{Metric Collection}          & \multicolumn{1}{c|}{\xmark}                  & \multicolumn{1}{c|}{loss, per iteration time} & \multicolumn{1}{c|}{\begin{tabular}[c]{@{}c@{}}finish-time \\ fairness estimate\end{tabular}} & \multicolumn{1}{c|}{per iteration time}      & \multicolumn{1}{c|}{loss, per iteration time} & \multicolumn{1}{c}{\begin{tabular}[c]{@{}c@{}}per iteration time, \\ resource utilization\end{tabular}} \\ \bottomrule
\end{tabular}}
\vspace{-10pt}
\end{table*}

\system is designed using the insight that almost all DL schedulers are created using a subset of the seven key abstractions demonstrated in Figure~\ref{fig:sched_schem}. \system's goal is to provide well defined API's for these abstractions and the ability to compose these abstractions to build a DL scheduler. Further, \system should facilitate creation of new abstractions and new instances of existing abstractions.
We first give a high level overview of how to use \system by showing an implementation for a simplified scheduler workflow. %as shown in Figure~\ref{fig:sched_schem}.

\system provides  a well defined API (detailed in \S~\ref{sec:api}) for all the abstractions described in Figure~\ref{fig:sched_schem} which are needed to build a scheduler.
The \emph{job admission policy} acts as a gate keeper for newly arriving jobs.
Each scheduling round, accepted jobs are queued to be scheduled on the cluster and the \emph{job scheduling policy} prioritizes a subset of all queued jobs to receive scheduling allocations that round.
A \emph{job placement policy} determines which server and specifically which of the accelerators on the server are assigned to each job that gets scheduled.
The \emph{job preemption} abstraction is responsible for preempting running jobs from the prior round which are not scheduled to run this round, or jobs whose placement has changed, by bundling up their state for subsequent launches or movement.
The \emph{job launch} abstraction is responsible to start new jobs for the round, or those that have moved, on destination servers.
Concurrently, a \emph{cluster manager} service constantly keeps track of job and cluster resource churn, and a \emph{metrics collector} aids in aggregating server-centric and job-centric statistics for use by other scheduler abstractions.

Table~\ref{tab:abstraction_used} describes the different instances of the abstractions needed by popular DL schedulers.

Figure~\ref{fig:blox_chain_flow} shows the implementation of a scheduler using \system in Python. Lines~\ref{line:ap} to~\ref{line:pp} create the job admission (\texttt{AcceptAll}), scheduling (\texttt{FIFO}) and placement (\texttt{Consolidated}) policy to use in our scheduler. Following that, we instantiate the \texttt{BloxManager}, a class that maintains endpoints for users to submit jobs and to communicate with workers. Next, we instantiate shared data structures that track the state of active jobs (\texttt{JobState}) and the state of active machines (\texttt{ClusterState}) in line~\ref{line:js} and~\ref{line:cs}. These data structures maintain the necessary shared state that can be used across modules and enable composition inside the scheduling loop (lines~\ref{line:sc-start} to~\ref{line:sc-end}). The scheduling loop contains the steps that are performed at every round of scheduling which we describe next.

At every round of scheduling, we first update the \texttt{ClusterState} to reflect any machines which have been added / removed (\texttt{update\_cluster}) and also update metrics of currently running jobs. We next prune any completed jobs and these three steps update our shared datastructures with progress from the previous round on all workers.

Following that, we retrieve new jobs which have been submitted for scheduling (\texttt{pop\_wait\_queue)} since the last round and invoke the acceptance policy (Line~\ref{line:accept}) to determine which of these new jobs should be accepted for scheduling.
The accepted jobs are added to \texttt{JobState}.
Having determined the set of schedulable jobs, we next invoke the scheduling policy (Line~\ref{line:schedule}) and pass relevant information necessary for scheduling through cluster and job states.
The scheduling policy returns a prioritized list of jobs that will be scheduled in this round, and we pass this list to the placement policy to determine which jobs should be executed on which GPUs.
The placement policy also determines which jobs, active in the prior round, should now be suspended.
Our final step in the scheduling loop is to pass in the list of jobs to be suspended and the list of jobs to be launched to the \texttt{BloxManager} (Line~\ref{line:exec}); job movement across two consecutive rounds effectively results in a suspension followed by a launch at its newly assigned placement.
The \texttt{BloxManager} coordinates with workers to preempt jobs that need to be suspended and renews the lease for jobs which will continue to run on the same workers (more details in Section~\ref{sec:eval}).
Overall, the above workflow shows an example of how developers can compose modules to create an end-to-end scheduler.

A workflow in \system{} can be used to deploy a scheduler on a cluster or to perform evaluations in simulation. As seen in lines~\ref{line:pop} and~\ref{line:sleep}, the developer only needs to set a command line argument to specify that this workflow run in simulation.

Further, with our modular design, the core logic of the scheduling workflow (\ie admission, scheduling and placement policies) remains same across simulation and cluster execution; this enables maximal code reuse across simulation and deployments, with the simulator skipping or using skeletal implementations of cluster management and job launch/preemption.

We discuss  design and implementation of \system in detail in \S~\ref{sec:design}.
In the next couple of sections though, we first discuss how to implement existing schedulers (\S~\ref{case-study:repro}), test them with evolving workloads (\S~\ref{case-study:comparison}) and deployments (\S~\ref{case-study:consolidation}). Finally we give examples of how new policies can be added and evaluated in  \system (\S~\ref{case-study:newsched}).

\section{Reproducing and Revisiting Schedulers}

\label{sec:case-studies}
In this section, we present case studies to highlight how \system{} can be used to build, compare, and understand existing DL schedulers (revisiting the past in new light).
Specifically, we focus on three case studies:
\begin{myitemizeleft}
\vspace{-5pt}
    \item We implement \emph{seven} existing DL schedulers in \sysname{} and validate accuracy of our implementation by reproducing some of their reported experiments (\S\ref{case-study:repro}).

    \item Study the performance properties (average JCT and responsiveness) of existing schedulers for new scenarios: (i) different workload traces (ii) varying cluster load to a point where resource contention is high (\S\ref{case-study:comparison}). % couple of
    \item Study the affect of placement preference on workloads due to changes in deployments and  evolution of workloads. We also study how using a profiling based approach can be more robust to these changes than fixed heuristics (\S\ref{case-study:consolidation}).
\end{myitemizeleft}

\paragraph{Workloads}
To evaluate existing policies, in this section, we use three different workloads traces.
Each workload trace contains a stream of job submissions with their arrival times, their requested number of GPUs, job execution duration (when run to completion in isolation).
In our experiments when we map a job to a particular workload (DNN model), we associate it with appropriate profile data such as its per-iteration time across different batch sizes and GPU count.
Unless otherwise specified, in this section our clusters are sized to have 128 GPUs, with each server having $4\times$ V100 GPUs (similar to Amazon EC2 \emph{p3.8xlarge}). Further all our experiments in this section are simulations, which is similar to prior work which use simulations to evaluate the schedulers~\cite{qiao2021pollux, gu2019tiresias, mahajan2020themis, mohan2021synergy}.
We verify the fidility of our simulation in Section~\ref{sec:eval}.
\begin{myitemizeleft}
\vspace{-4pt}
 \item \phtrace{}: We use the production traces derived from Microsoft's Philly Cluster~\cite{jeon2019analysis}.
 Similar to prior work~\cite{mahajan2020themis, narayanan2020heterogeneity, mohan2021synergy, qiao2021pollux}, we randomly assign jobs to use one of the models listed in Table~\ref{tab:models} to each job. To vary load in the cluster we assign job arrival times using a Poisson arrival process with the inter-arrival rate of $\lambda$. Varying $\lambda$ modifies the job arrival rate, allowing us to generate different amounts of the load. Similar to prior work~\cite{narayanan2020heterogeneity, mohan2021synergy}, in simulation, we track the progress of jobs with ID 3000 to 4000 in the trace and use their completion times to compute average JCT. This ensure we study steady state behavior with new jobs continuing to arrive until jobs of interest complete.
 \item \potrace{}: We use the trace which was open sourced  by  the authors of Pollux~\cite{polluxtrace}. The trace contains 160 jobs samples from the busiest 8 hour window from the Microsoft trace~\cite{jeon2019analysis} and we use this to study the behavior of Pollux. More details on this trace can be found in \cite{qiao2021pollux}.
 \item \titrace{}: To reproduce the results in Tiresias we use the trace used in their paper; csv-60 from their open source code repository~\cite{opensourcetiresias}.
 \end{myitemizeleft}
%\shivaram{Add how do we compute Average JCT.}

\begin{table}[t]
\caption{\small{Models used in \system{} to evaluate schedulers using \phtrace{}}}
\label{tab:models}
\vspace{-10pt}
\resizebox{\linewidth}{!}{
\begin{tabular}{@{}lll@{}}
\toprule
Model Name  & Dataset     & Task                          \\ \midrule
Resnet-18~\cite{he2016deep}   & Cifar-10    & Image Classification          \\
CycleGan~\cite{zhu2017unpaired}    & monet2photo & Image to Image Transformation \\
Resnet-50~\cite{he2016deep}   & Imagenet    & Image Classification          \\
LSTM~\cite{hochreiter1997long}        & WikiText-2  & Next word prediction          \\
Recoder~\cite{moussawi2018towards}     & ML-20M      & Recommendation                \\
Transformer~\cite{vaswani2017attention} & Multi30K    & Language Translation          \\
A3C~\cite{mnih2016asynchronous}        & Pong        & Deep RL                       \\ \bottomrule
\end{tabular}}
\end{table}

\subsection{Reproducing existing DL schedulers}
\label{case-study:repro}
\begin{table}[t]
\vspace{-10pt}
\caption{\small{Modules and the number of lines of code added to implement specific schedulers in \sysname{}}}
\vspace{-10pt}
\label{tab:modules_changed}
\resizebox{\linewidth}{!}{
\begin{tabular}{@{}l|l|l@{}}
\toprule
Scheduler Name & Abstractions modified                                                                                                                                           & Lines of Code \\ \midrule
LAS            & Scheduling Policy                                                                                                                                      & 12            \\
Tiresias       & Scheduling Policy, Placement Policy                                                                                                                    & 295           \\
Optimus        & Scheduling Policy, Metric Collection, Placement Policy                                                                                                 & 246           \\
Gavel          & Scheduling Policy, Metric Collection, Placement Policy                                                                                                 & 539           \\
Pollux         & Scheduling Policy, Metric Collection, Workload Generation                                                                                              & 1157          \\
Themis         & Scheduling Policy, Metric Collection                                                                                                                   & 745           \\
Synergy        & Scheduling Policy, Placement Policy, Workload Generation & 1137 \\ \bottomrule
\end{tabular}}

\end{table}
\begin{figure}[t]
    \centering
    \includegraphics[width=0.9\linewidth]{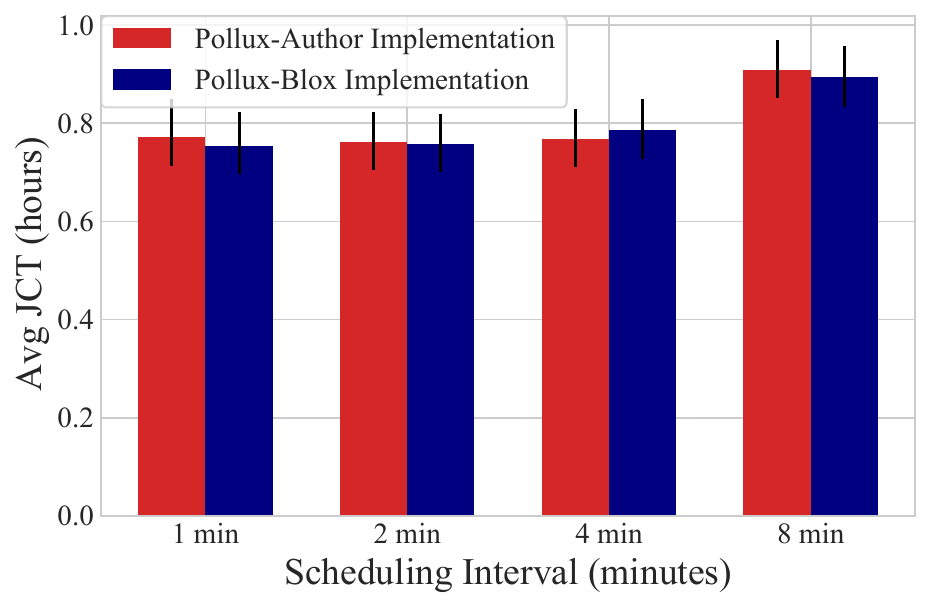}
    \vspace{-15pt}
    \caption{\small{\textbf{Reproducing Pollux.} We reproduce the experiment in Section 5.3.2 from the Pollux paper~\cite{qiao2021pollux}
    using the Pollux implementation in \sysname{}.}}
    \label{fig:pollux_verification}
    \vspace{-10pt}
\end{figure}

\begin{figure}[t]
    \centering
    \vspace{-5pt}
    \includegraphics[width=0.9\linewidth]{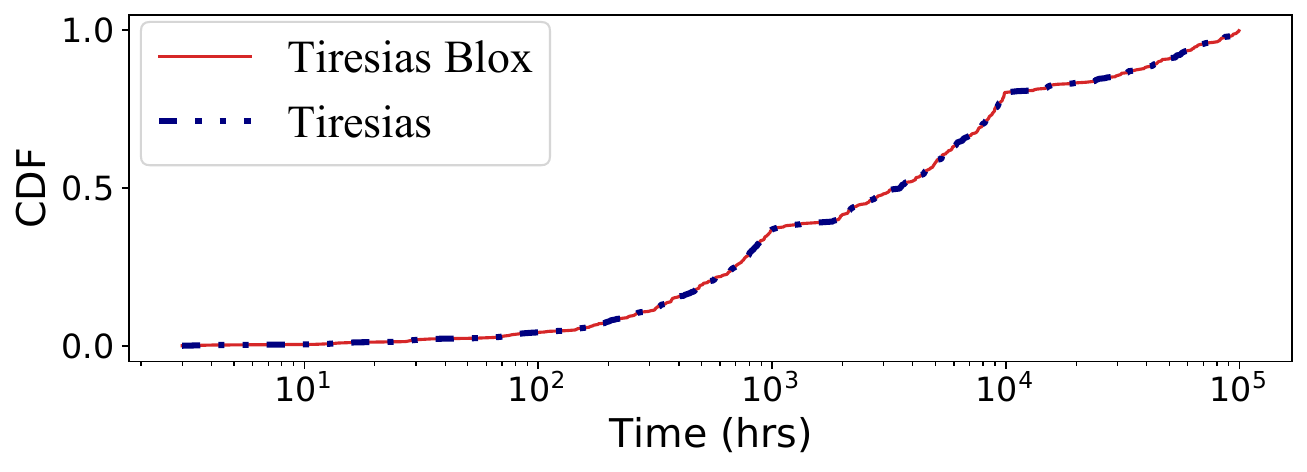}
    \vspace{-10pt}
    \caption{\small{\textbf{Reproducing Tiresias.} Comparing open source Tiresias with its implementation in \system{} when run on \titrace{}.}}
    \label{fig:tiresias_verification}
    \vspace{-20pt}
\end{figure}
We first demonstrate the flexibility of \system by implementing a number of existing schedulers that have been developed in prior work. We have implemented the following \emph{seven} schedulers (Table~\ref{tab:abstraction_used}): First in First Out (FIFO) used in many prior schedulers including Philly~\cite{jeon2019analysis}, single-queue Least Attained Service (LAS) and discreet-LAS from Tiresias~\cite{gu2019tiresias}, Optimus~\cite{peng2018optimus}, heterogeneity-aware LAS from Gavel~\cite{narayanan2020heterogeneity}, Pollux~\cite{qiao2021pollux}, Finish Time Fairness (FTF) from Themis~\cite{mahajan2020themis}, and Synergy~\cite{mohan2021synergy} in \system{}.
To estimate the implementation overhead for each of these prior frameworks, we start with a FIFO scheduler as the baseline and then count the number of modules that need to be updated or added to realize a particular system.

Table~\ref{tab:modules_changed} lists the modules and the number of lines of code required to implement these seven DL scheduling frameworks. We see that most schedulers require changing two or three modules and  a relatively small number of lines of code change (100s). The two exceptions here are Pollux and Synergy. Pollux includes code to evaluate training efficiency based on convergence and optimize for goodput~\cite{qiao2021pollux} and uses a workload trace with a different schema. So we had to add a new workload parser resulting in around 350 extra lines of code. Synergy proposes a number of placement strategies including an optimization-based strategy that required around 500 lines of code. Overall, our results demonstrate that users can implement a wide variety of DL schedulers in \system with minimal changes.

\paragraph{Verfiying Existing Scheduler Implementations} We next verify that our implementations of the aforementioned schedulers are faithful by reproducing experiments from three prior works: Pollux~\cite{polluxrepo}, Synergy~\cite{mohan2021synergy} and Tiresias~\cite{gu2019tiresias}.
To meaningfully compare experimental results of the \sysname{} implementation of these schedulers to those of the baseline systems, we use cluster sizes and workload profile data (such as a workload's per-iteration time) as specified in the original experiments for the respective schedulers.
For Pollux, we use \potrace{} and reproduce the experiment in Section 5.3.2 from the Pollux OSDI 2021 paper~\cite{qiao2021pollux}.

\begin{figure}[t]
    \begin{center}
    \begin{subfigure}[b]{0.47\linewidth}
    \includegraphics[width=0.9\linewidth]{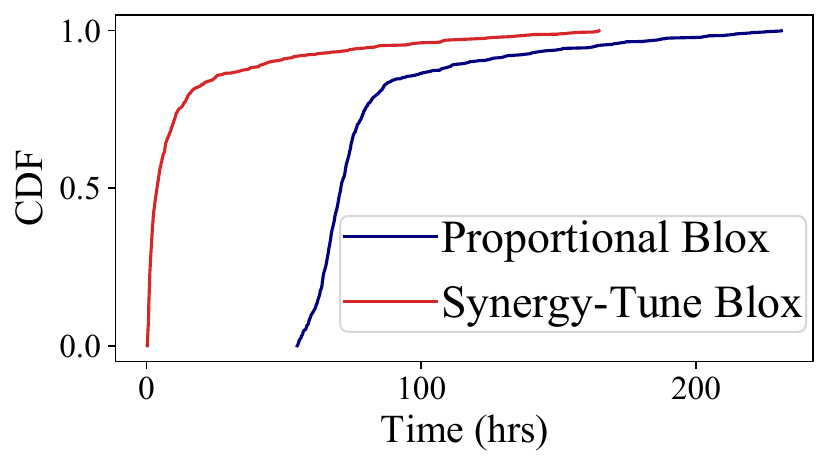}
    \vspace{-5pt}
    \caption{Synergy in \system. }
    \end{subfigure}
    \begin{subfigure}[b]{0.47\linewidth}
    \includegraphics[width=0.9\linewidth]{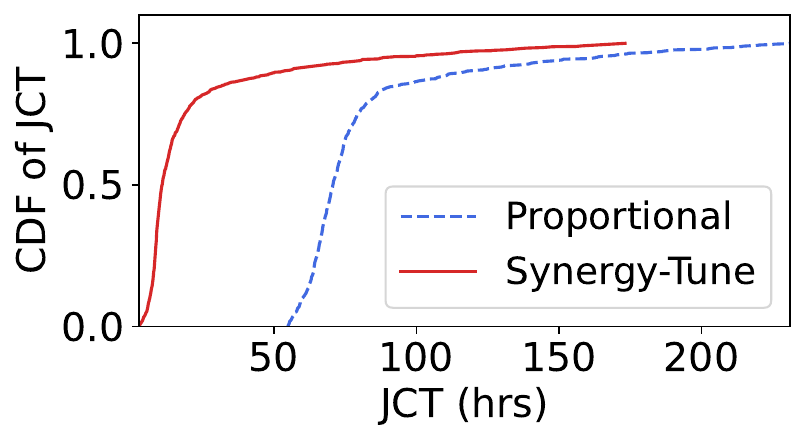}
    \vspace{-5pt}
    \caption{Extracted from Synergy logs}
    \end{subfigure}
    \vspace{-10pt}
    \caption{\small{\textbf{Reproducing Synergy.} Synergy's Proportional and Tune policies in \sysname{} (left) match the original Synergy implementation (right, and Figure 9(b) from that  paper~\cite{mohan2021synergy}).}}
    \label{fig:synergy_verification_cdf}
    \end{center}
\vspace{-15pt}
\end{figure}

We measure the average job completion time while varying the scheduling interval (scheduling round duration).
Figure~\ref{fig:pollux_verification} shows that results from \system{} closely match the Pollux open source implementation (maximum deviation of 2.4\%) and we also verify that these numbers closely match those reported in the Pollux paper~\cite{qiao2021pollux}.
For Tiresias~\cite{gu2019tiresias}, Figure~\ref{fig:tiresias_verification} similarly shows that our implementation in \system{} matches the Tiresias open source simulator when we measure the CDF of JCTs while run with \titrace{}.
Finally, Figure~\ref{fig:synergy_verification_cdf} shows that we can also accurately reproduce Figure 9(b) from the Synergy OSDI 2022 paper~\cite{mohan2021synergy} and find that \system{} exactly matches the  CDF of JCTs for both modes (Proportional, Synergy-Tune) when run with the \phtrace{}.

\noindent\textit{Takeaway: We are able implement  a wide variety of DL schedulers in \sysname{} with relatively minimal code changes and are able to accurately reproduce results from a number of prior scheduling frameworks.}

\subsection{Comparing scheduling policies}
\vspace{-5pt}
\begin{figure}[t]
    \centering
    \includegraphics[width=0.9\linewidth]{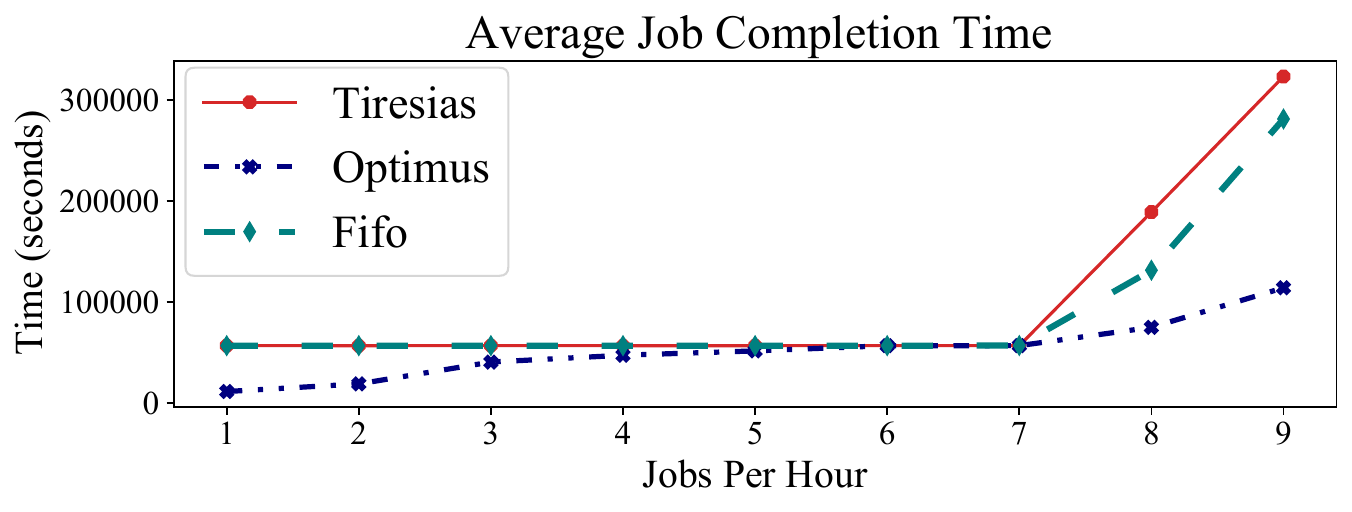}
    \vspace{-10pt}
        \caption{\small{\textbf{Scheduling Policies JCT.} Comparing FIFO, Tiresias, and Optimus on \phtrace{} for varying loads (1 to 9 jobs/hour.}}
    \label{fig:jct_optimus_tiresias}
    \vspace{-10pt}
\end{figure}

\begin{figure}[t]
    \centering
    \includegraphics[width=0.9\linewidth]{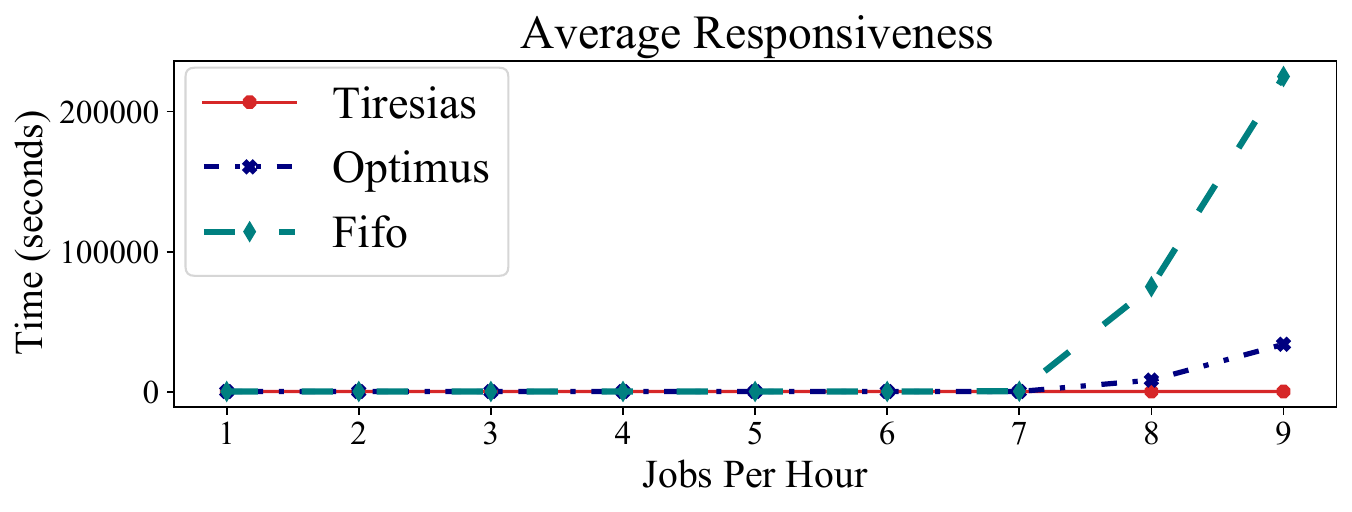}
    \vspace{-10pt}
    \caption{\small{\textbf{Scheduling Policies Responsiveness.} Comparing FIFO, Tiresias, and Optimus on \phtrace{} as we vary load from 1 job/hour to 9 jobs/hour.}}
    \label{fig:responsiveness_optimus}
    \vspace{-15pt}
\end{figure}

\label{case-study:comparison}
Having different schedulers implemented in the same system allows us to perform a fair comparison between existing scheduling policies across different metrics while varying the load.
For these experiments we use the \phtrace{} and vary the job arrival rate from 1 job/hour to 9 jobs/hour.
We use two metrics: average job completion time and responsiveness.   While average job completion time (JCT) is a well studied metric, we also study the trade-offs with respect to \textit{responsiveness}. Responsiveness for a job is defined as the time elapsed between when the job was received by the scheduler and when the job was first scheduled. Responsiveness can also be interpreted as the time taken for a user to get first feedback on a job. For both average JCT and responsiveness, a lower value is desirable.

We compare three different scheduling policies: FIFO, Tiresias and Optimus in Figures~\ref{fig:jct_optimus_tiresias} and~\ref{fig:responsiveness_optimus}, and use consolidated placement for all policies.
From the figure we see that at low load ($<$ 4 jobs/hour), Optimus has a lower average JCT than FIFO and Tiresias but with similar responsiveness; this is because Optimus assigns more resources to jobs closer to completion (convergence).
At higher loads ($>$ 7 jobs/hour), we observe a different behavior: Tiresias has a higher JCT than FIFO and Optimus.
Since Tiresias gives newly arriving jobs a shot at receiving early allocations and prioritizes jobs with least attained service, leading to improved responsiveness, it also causes long running jobs to suffer a large number of preemptions thus having longer average JCT at high load.
On the other hand Optimus prioritizes jobs which will converge faster thus leading to lower average JCT, but sacrifices responsiveness (compared to Tiresias).
As expected, FIFO has the worst responsiveness under high load.

\begin{figure}[t]
    \centering
    \includegraphics[width=0.9\linewidth]{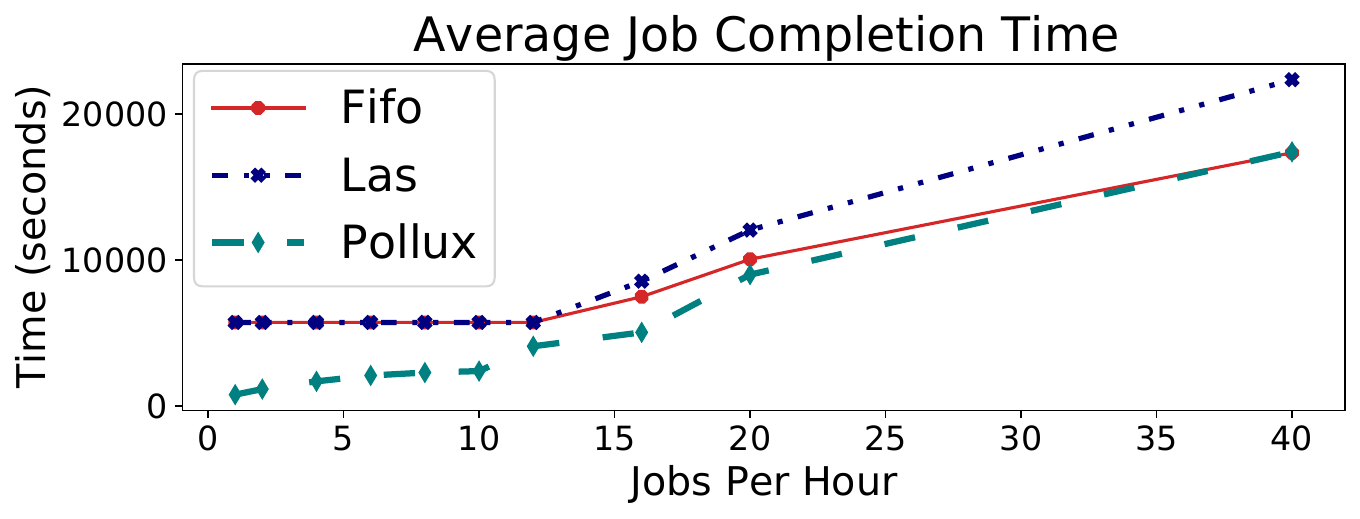}
    \vspace{-10pt}
    \caption{\small{\textbf{Scheduling Policies JCT.} Comparing Pollux, FIFO and simplified single-queue LAS on the \potrace{} using 64 GPUs as we vary load from 1 job/hour to 40 jobs/hour.}}
    \label{fig:jct_pollux}
    \vspace{-10pt}
\end{figure}

\begin{figure}[t]
    \centering
    \includegraphics[width=0.9\linewidth]{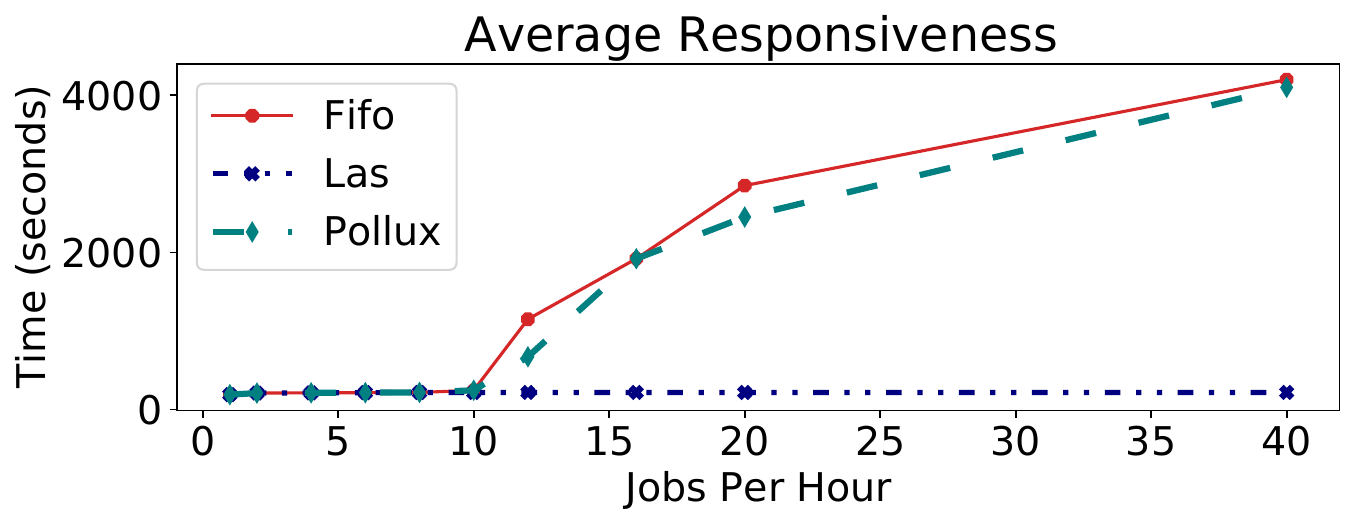}
    \vspace{-10pt}
    \caption{\small{\textbf{Scheduling Policies Responsiveness.} Comparing Pollux, FIFO and simplified single-queue LAS on the \potrace{} using 64 GPUs as we vary load from 1 job/hour to 40 jobs/hour.}}

    \label{fig:responsiveness_pollux}
    \vspace{-10pt}
\end{figure}

To study the JCT and responsiveness trade-off for Pollux, we repeat the same experiment using the \potrace{} as that has the necessary batch size and convergence information used by the Pollux scheduler.
In Figures~\ref{fig:jct_pollux} and~\ref{fig:responsiveness_pollux}, we compare Pollux against FIFO and single-queue LAS scheduling policies  while using consolidated placement.
We increase the load (in terms of jobs/hour) to a larger number than in Figures~\ref{fig:jct_optimus_tiresias} and~\ref{fig:responsiveness_optimus} as the majority of jobs in \potrace{} have sub-10-hour runtimes (when run to completion in isolation), and this mandates a higher load for resource contention to kick in compared to \phtrace{}.
From the figures we can see that at low to medium load (under 15 jobs/hour), Pollux offers improvements in average JCT compared to the other two policies while being equally responsive.
This is because Pollux can dynamically change the batch size and number of GPUs used by jobs when there are enough resources available.
However, as load increases we see that Pollux's responsiveness and JCT become similar to FIFO (> 20 jobs / hour).
Our analysis indicates,that this happens due to contention for resources increases at high load. Pollux, which avoids job preemptions, allocates fewer GPUs to running and incoming jobs (a single GPU at high loads) instead of their actual GPU demand with the goal of increasing goodput. % instead
%as resources become available later.  allocation
However, at sufficiently high load, if there are more jobs than GPUs available, the incoming jobs are queued affecting responsiveness.
Finally, we also see that LAS maintains good responsiveness even at high load because it preempts long running jobs and offer resources to incoming jobs.

\noindent\textit{Takeaway: With \system{}, we can study the trade-offs in existing schedulers under varying load, and observe interesting properties. At high load: FIFO can have lower JCT than Tiresias while sacrificing responsiveness also the performance of Pollux degrades becoming similar to FIFO.} %\shivaram{Amar: better phrasing?}

\begin{figure}[t]
    \centering
    \includegraphics[width=0.9\linewidth]{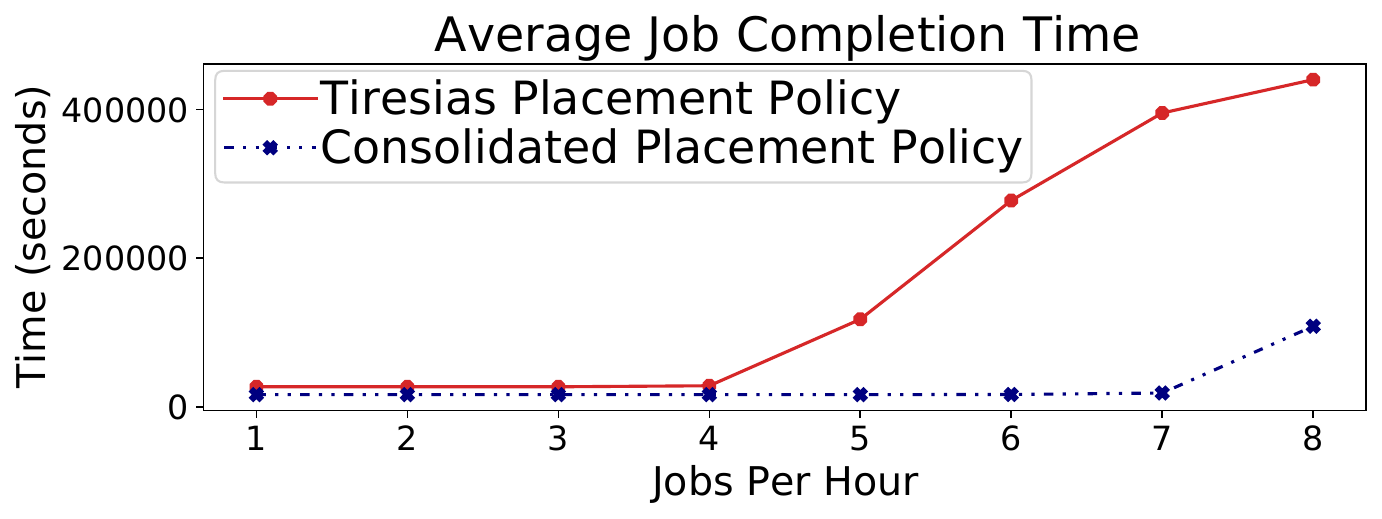}
    \vspace{-10pt}
    \caption{\small{\textbf{Placement policies on V100:} Comparing average JCT with \phtrace{} for Tiresias placement policy vs. a placement policy that consolidates all jobs. Lower bandwidth (10 Gbps) and faster compute (V100 GPUs) leads to consolidation performing better at high load.}}
    \label{fig:placement-v100}
    \vspace{-10pt}
\end{figure}

\subsection{Revisiting Placement Policies}
\label{case-study:consolidation}
\system{} provides us the ability to study how changes in hardware or workload can affect design decisions made in DL schedulers. With the rapid deployment of new DL-specific hardware (e.g., A100 GPUs, TPUs, GraphCore etc.), the balance between computation and communication in model training is continuously evolving. Similarly, the models that are being trained on enterprise clusters are also evolving, from CNN models such as VGG19, AlexNet to Transformer-based models such as BERT~\cite{devlin2018bert} and GPT-3. Thus, placement policies that determine where jobs are placed in the cluster need to be re-evaluated due to hardware and workload changes.

\paragraph{Varying cluster setup.} To study the above scenario, we consider the placement policy proposed in Tiresias. The Tiresias placement policy selectively performs consolidation only for jobs which have a high degree of skew across tensors in a model (Section 3.3 in~\cite{gu2019tiresias}), and remaining jobs are placed to minimize fragmentation.
The authors show that this policy can improve overall JCT on a cluster of servers with 4xP100 GPUs, with 100Gbps interconnect across machines.
We revisit this experiment using the \phtrace{}, but with a cluster of servers with 4xV100 GPUs on AWS (p3.8xlarge machines) which have more computation power but only have a 10Gbps interconnect across servers.
Figure~\ref{fig:placement-v100} compares the average JCT while varying load when using the Tiresias placement policy to a policy that consolidates placement for all jobs.
From the figure we can see that on the V100 cluster, the consolidated placement policy performs better at higher loads (greater than 4 jobs/hour).
This is because the V100 cluster has higher computation power and a worse network interconnect than what was in the private cluster used in the initial Tiresias study, making it more likely that communication is a bottleneck for model training, hence favoring consolidation for all models.
Thus, we see that the placement policies need to be guided by profiles on specific hardware they are deployed on, rather than using fixed heuristics.

\begin{figure}[t]
    \centering
        \includegraphics[width=0.9\linewidth]{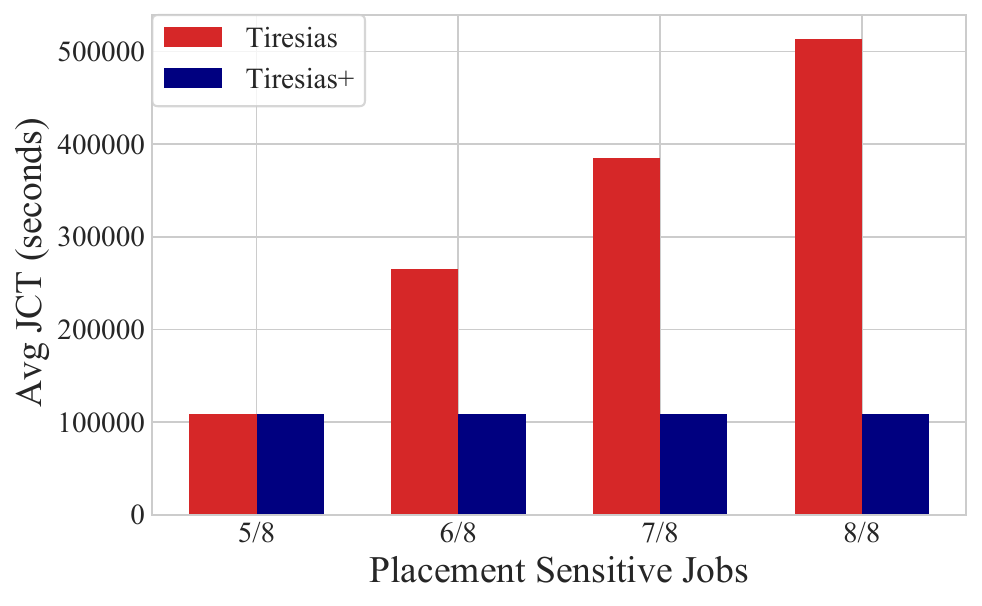}
        \vspace{-10pt}
    \caption{\small{\textbf{Placement policies with profiles:} Average JCT as we vary the number of placement sensitive jobs in the \phtrace{}. We compare the placement policy from Tiresias to a policy that has perfect knowledge of which workloads are placement sensitive.}}
    \label{fig:workload_mix}
    \vspace{-10pt}
\end{figure}

\paragraph{Varying model properties} We next consider how varying the workload mix in terms of model properties can affect placement policies.
To study this, we consider the same Tiresias setup on the V100 cluster as in the previous section, for a load of 8 jobs/hour, but change our workload mix in the trace such that initially there are only 5 out of 8 workloads that benefit from placement consolidation.
We compare two policies with this setup: the baseline Tiresias placement policy that uses the skew-based consolidation heuristic~\cite{gu2019tiresias} and Tiresias+ which uses a placement policy that has perfect knowledge of which models benefit from consolidation (can be realized with profiled data).
Both these policies respect the idea introduced by Tiresias that a distributed DL job that does not benefit from consolidation on the same machine can be safely fragmented across servers.
We then incrementally increase the number of workloads that prefer consolidation until 8; the skew-based heuristic in the baseline scheme is only able to identify the first 5 workloads as benefiting from placement consolidation.
Figure~\ref{fig:workload_mix} shows the average JCT for these policies and we find that Tiresias+ has the lowest average JCT, and the gap between the baseline placement policy and profile-based placement policy grows as we increase the number of workloads that benefit from consolidation, thus highlighting the benefits of having accurate profiles as workloads evolve to guide placement decisions.

\noindent\textit{Takeaway: Placement policies that consider consolidation preferences of jobs are a good idea. But the placement preference of workloads can be affected by changes in deployments and evolution of workloads; using a profiling based approach is robust to these changes than fixed heuristics, and using \system{} we are easily able to compare and study this effect.}

\section{Designing New Schedulers with Blox}
\label{case-study:newsched}
In this section, we show how to realize new scheduler designs by composing modules in \system{}.  Specifically, we focus on:

\begin{myitemizeleft}
    \item Studying the effectiveness of new scheduler designs that can trade-off average JCT vs. responsiveness and how such designs can be realized easily by composing admission policies and scheduling policies in \system{} (\S\ref{case-study:composition}).
    \item Using \system to build an automatically synthesizing scheduler which based on job arrival patterns and job duration is able to automatically %choose the best instances of the abstractions
    compose new schedulers to optimize a operator preferred metric. (\S\ref{sec:automatic_scheduler})
    \item Studying the flexibility of \sysname{} in supporting the addition of new policies by highlighting the ease with which we can prototype and evaluate a new loss-based job termination and intra-node job placement policy  (\S\ref{case-study:loss-term}).
\end{myitemizeleft}

\subsection{Composing admission and scheduling}
\label{case-study:composition}

We study  composing and prototyping new policies, in the context of LAS where our previous experiments in Figures~\ref{fig:jct_pollux} and~\ref{fig:responsiveness_pollux} showed that the average JCT can increase significantly at high loads. Here we investigate if adding an admission policy that restricts the set of schedulable jobs can improve JCT while sacrificing some responsiveness.

\paragraph{FIFO Admission Control with LAS scheduling} To realize the above idea in \sysname{}, we compose a FIFO admission block with the LAS scheduling and consolidated placement blocks. We perform admission control as follows: once the number of GPUs requested by admitted jobs (i.e., schedulable jobs) crosses a threshold (e.g., $1.5x $ the number of GPUs available in the cluster), we enqueue newly arriving jobs in the admission control block. Jobs are released for scheduling in a FIFO manner as resources become available. Once jobs have been admitted, they are scheduled using the same LAS policy.

\begin{figure*}[t]
    \centering
    % \vspace{-10pt}
    \begin{minipage}[t]{0.3\linewidth}
    \includegraphics[width=\linewidth]{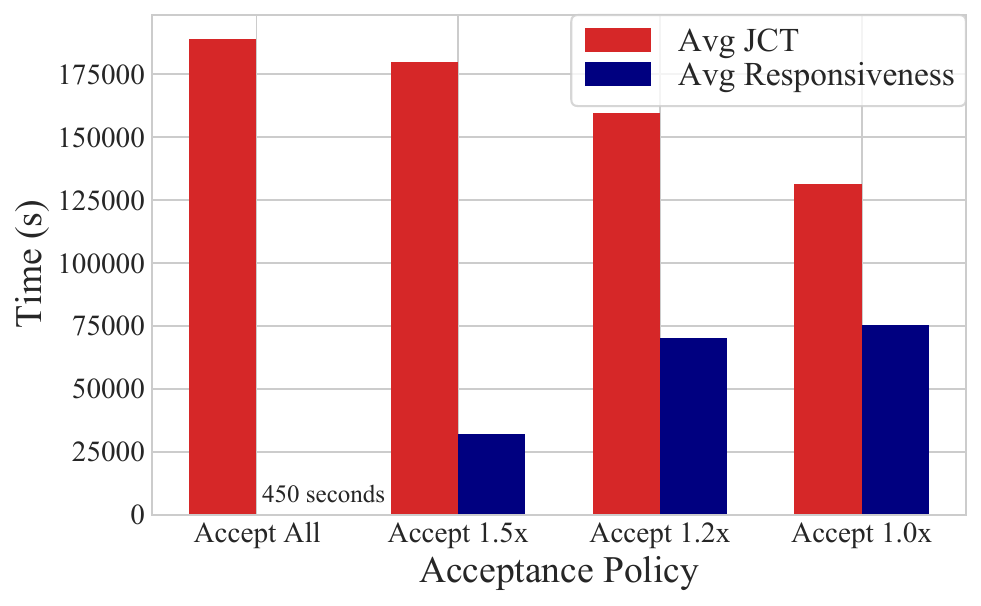}
    \vspace{-15pt}
    \caption{\small{\textbf{Composing policies:} We use a FIFO admission policy with LAS scheduling policy, we can trade-off between JCT and responsiveness. Accept $1.5\times$ is 5\% faster, Accept $1.2\times$ is 15\% faster and Accept $1.0\times$ 30\% faster for the \phtrace{} with 8 jobs/hr.}}
    \label{fig:acceptance_policy_standard_workload}
     \vspace{-10pt}
     \end{minipage}\quad
     \begin{minipage}[t]{0.3\linewidth}
    \includegraphics[width=\linewidth]{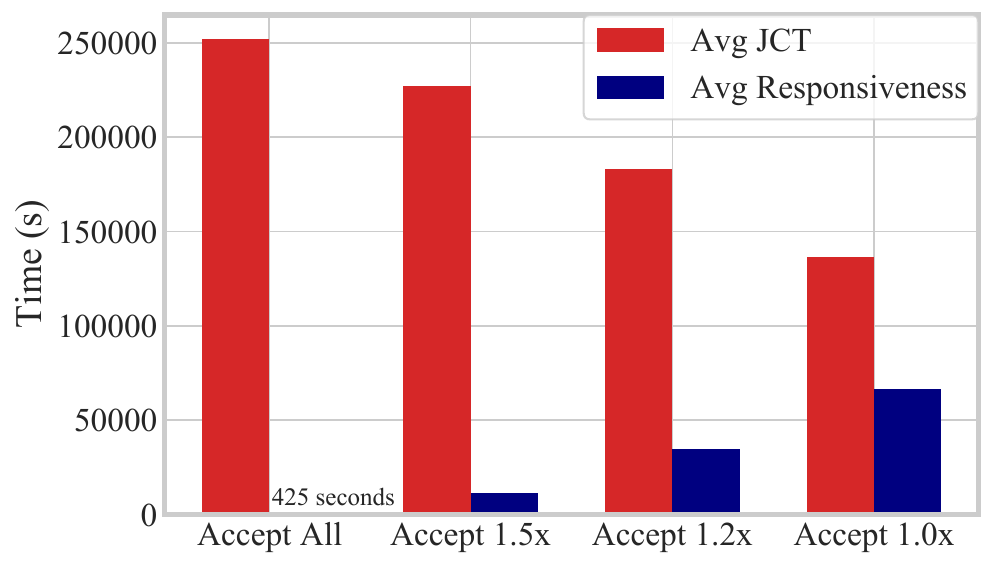}
    \vspace{-15pt}
    % \vspace{-5pt}
    \caption{\small{\textbf{Handling Workload Spikes:} Using a FIFO admission policy with LAS while varying the admission control threshold. Using \phtrace{} with 8 jobs/hr and a spike of 16 jobs in one hour each day, Accept $1.2\times$ has 27.3\% lower JCT than Accept All.}}
    \label{fig:acceptance_policy_spiky}
    \vspace{-10pt}
     \end{minipage}\quad
     \begin{minipage}[t]{0.3\linewidth}
         \includegraphics[width=\linewidth]{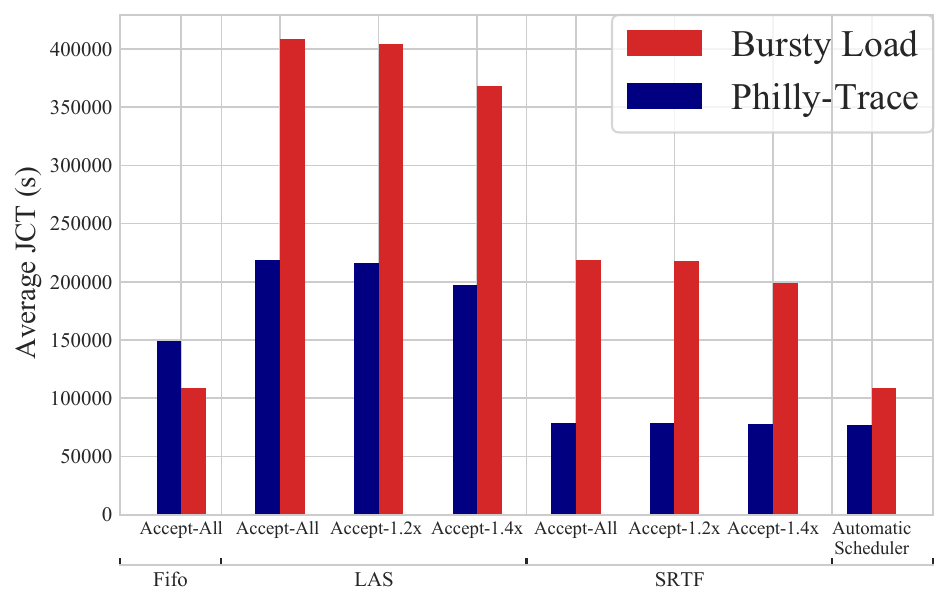}
    \vspace{-15pt}
    \caption{\small{\textbf{Average JCT's of Automatic Scheduler:} Automatic scheduler closely matches the performance of the best performing static policy for both \phtrace{} and Bursty Load. }}
    \label{fig:dynamic_policy_jct}
    \vspace{-10pt}
     \end{minipage}

\end{figure*}

We next compare how varying the acceptance threshold affects JCT and responsiveness using the \phtrace{} with an arrival rate of 8 jobs/hour.
Figure~\ref{fig:acceptance_policy_standard_workload} shows that composing an admission policy is able improve average JCT (by 15\% with Accept 1.2x) but that this can lead to worse responsiveness (up to 46\% of average JCT).  We also study if admission control can further help in a scenario where we have a sudden spike of job arrivals: using the same \phtrace{} with an arrival rate of 8 jobs/hour, we inject an additional 16 jobs during one hour in each day. We find that using an acceptance policy along with LAS leads to further benefits in this scenario (Figure~\ref{fig:acceptance_policy_spiky}), with average JCT improving by 15.4\% with Accept 1.5x and 27.3\% with Accept 1.2x.

\noindent\textit{Takeaway: By composing different modules in \system{} we are able to easily realize new schedulers. In this scenario, we see that using admission control policies along with LAS is one effective way to trade responsiveness for improvements in average JCT.}

\subsection{Automatically Synthesizing Schedulers}
\label{sec:automatic_scheduler}
As observed in \S~\ref{sec:case-studies}, different schedulers behave differently under varying loads (\S~\ref{case-study:comparison}), workload composition (\S~\ref{case-study:consolidation}) and cluster setup (\S~\ref{case-study:consolidation}). In real world setups the load average can be highly variable. Further in \S~\ref{case-study:composition} we show that different combination of admission polices can also affect the JCT of the jobs. However, existing schedulers are usually designed with a specific workload and metric in consideration, \eg Pollux is designed to improve throughput in medium load situations, while SRTF prioritizes short jobs to improve JCT. A cluster operator usually has to pick one of these policies based on experience and existing schedulers make it very hard to swap between different policies.

\system's{} modular design allows operators to easily swap different modules. Therefore, it is easy to combine different instances of abstractions, to compose a scheduler which optimizes for a given metric.
Using \system{} we build an automatic scheduler synthesizer which combines different abstractions (at runtime!) to improve a given metric.
To decide which instance of the available abstraction to run, every ten rounds (a round is five minutes) we run a simulation in parallel for all possible combinations with the same cluster setup and the available jobs on the cluster, \eg suppose there are two different admission policies and two different scheduling policies, we create all four possible combinations. We use this simulation to collect the metrics of importance and choose which combination of policies to run in order to maximize the metric of interest.

The goal of our automatic scheduler synthesizer is to choose the best possible combination of scheduling and job admission policy.
For our experiments we choose three scheduling policies- FIFO, SRTF and LAS - and three job admission policies - Accept All, Accept-$1.2\times$ and Accept-$1.4\times$. Accept All,  means all jobs are admitted into the cluster, Accept-$1.2\times$ and Accept-$1.4\times$ indicates that total cumulative resource requirements of all the jobs accepted to run are $1.2\times$ and $1.4\times$ of the GPU resources available on the cluster respectively.
 For evaluation we use \phtrace{}, and
 a bursty \phtrace derived workload (similar to one used in \S~\ref{case-study:composition}), where we send short bursty jobs at two times the load for two consecutive hours every four hours.
 For example, if the usual load is around eight jobs/hr, we send two times the load of short jobs (runtime chosen randomly  between ten minutes and one hour) for two consecutive hours after every four hours. This creates bursty load with a lot of short jobs.

\begin{figure}[t]
\begin{center}
    \begin{subfigure}[t]{\linewidth}
    \includegraphics[width=0.9\linewidth]{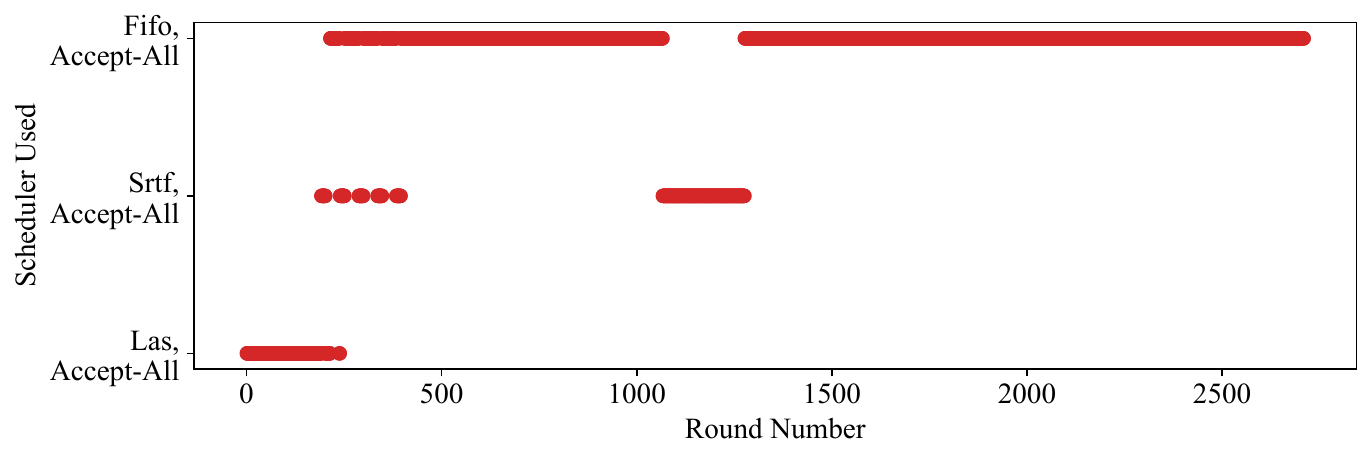}
    \vspace{-8pt}
    \caption{\small{Bursty Load}}
    \label{fig:scheduler_used_bursty}
    \end{subfigure}
    \end{center}

\begin{center}
    \begin{subfigure}[t]{\linewidth}
    \includegraphics[width=0.9\linewidth]{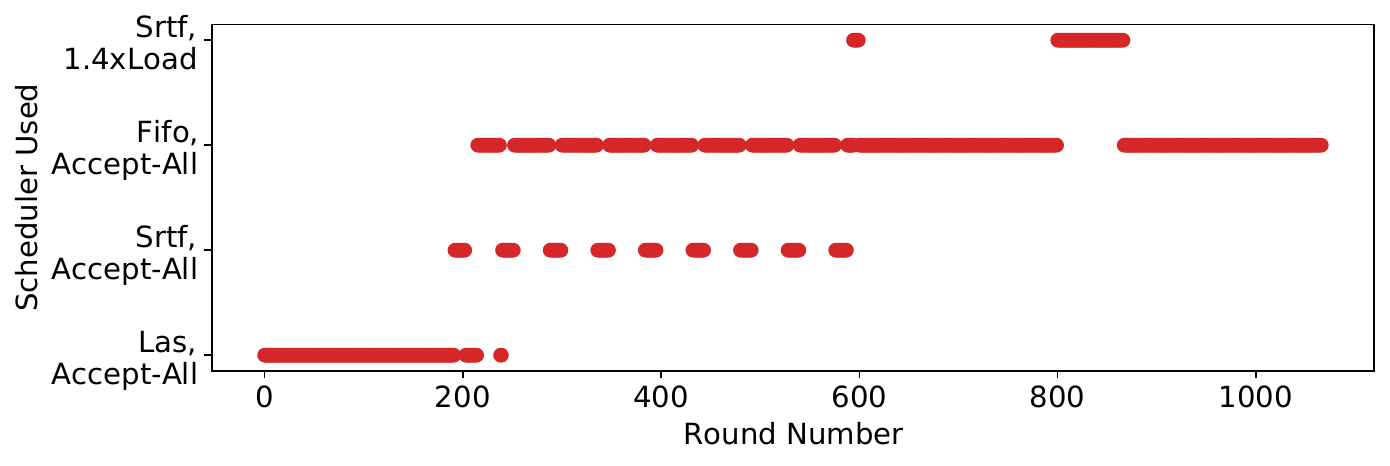}
    \vspace{-8pt}
    \caption{\small{\phtrace{}}}
    \label{fig:scheduler_use_original}
    \end{subfigure}
    \end{center}
\vspace{-12pt}
\caption{\small\textbf{Automatic Scheduler policies choice:} The temporal distribution of scheduler used by automatic scheduler for the bursty load and the \phtrace. We observe that dynamic policy is able to continuously switch among different scheduling policies.}
\label{fig:temporal_distribution}
\vspace{-10pt}
\end{figure}

Our goal using automatic scheduling synthesizer in this experiment is to improve average JCT.
In Figure~\ref{fig:dynamic_policy_jct} we compare JCT's for the two different workloads. For the \phtrace{} we observe that FIFO provides best average JCT's for jobs in range 3000-4000 while SRTF provides the best average JCT for bursty workload.
In Figure~\ref{fig:temporal_distribution} we show which scheduler was chosen by our Automatic Scheduler Synthesizer. In Figure~\ref{fig:temporal_distribution} we  observe that the choice of the best policy heavily depends on the trace and workload, and can not be determined apriori, thus necessitating an approach like ours.

In Appendix~\ref{appendix:additional_automatic} we show additional results how \system{} can be used to optimize multiple metrics like average JCT and responsiveness at the same time. In future we plan to extend this and rather than using simulation use a learning based approach to determine the policies to choose.

\begin{figure}[t]
    \includegraphics[width=0.9\linewidth]{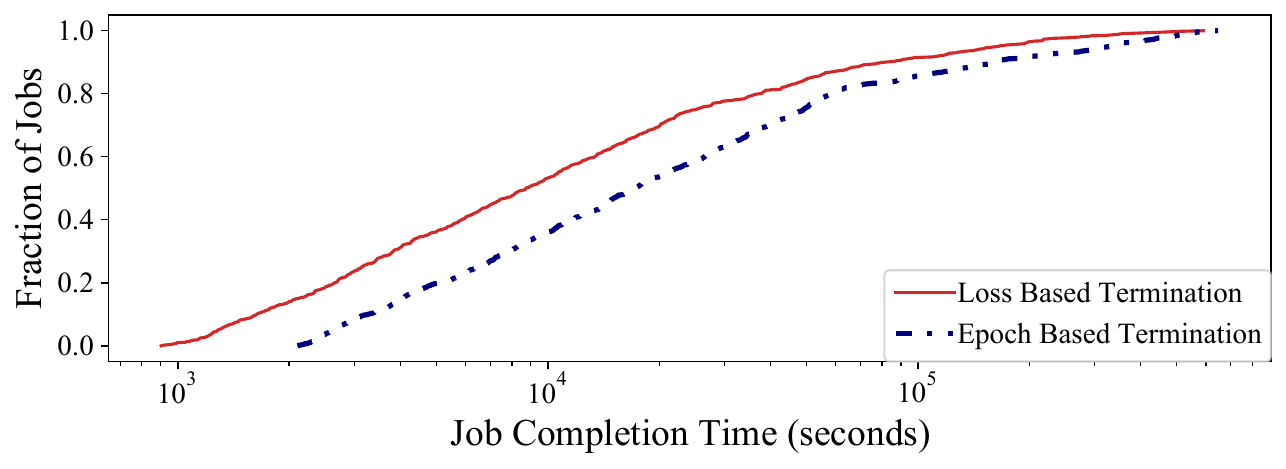}
    \centering
    \vspace{-10pt}
    \caption{\small{\textbf{Loss-based termination.} For a FIFO scheduling policy and \phtrace{} with 7 jobs/hour,  loss-based termination can reduce the Avg JCT by almost 44\%.}}
    \label{fig:loss_submis}
    \vspace{-20pt}
    \end{figure}

\subsection{Adding New Policies}
Next we give a couple of examples of adding completely new policies to \system{}.

\label{case-study:loss-term}
\paragraph{Supporting loss-based termination}

Prior work~\cite{jeon2019analysis} has observed that ``around 75\% of jobs reach within 0.1\% of lowest loss using only 40\% of the epochs''. This indicates that ML engineers typically overestimate the number of epochs needed to reach the desired loss value for their models.  To study the benefits of this observation, we add a new loss based job termination policy in \sysname{} with just \emph{4 additional lines of code}. The policy we implement is the following: for each job we take as input an additional parameter determining the relative loss threshold for termination (e.g., 0.2\%). Next, in the scheduling policy we add code to check if the current loss value for the job, collected by the \sysname{} \textit{Metric Collector}, is below the threshold and if this is the case, mark the job as completed and ready for termination.
The loss metric for each job is collected by the \cscheduler{} using the \blcl which provides an API to push any application specific metric. \system ensures that these metrics are available when the scheduler calls the \textit{Metric Collector}. %The \blcl provides an API, to allow applications to push any metric allowing

We evaluate our loss based termination policy using the \phtrace{}. Based on the observation in~\cite{jeon2019analysis},  we randomly assign 75\% of the jobs to converge in 40\% of their training time. Figure~\ref{fig:loss_submis} shows the CDF of job completion times when using loss-based termination policy vs the default epoch-based termination policy. Compared to using the number of epochs specified by the job, we observe that using loss-based termination leads to by around 44\% reduction in average JCT. We note that this result is from our simulation and we use a trace that contains per-job loss progression for this experiment; supporting loss-based termination in real-world deployments requires users knowing what the target accuracy for the model they are training ought to be (an insight some users might not be aware of).

\paragraph{Intra-Node Placement Policies}
Next, we show how to add additional placement constraints beyond just the regular placement policies as discussed in Section~\ref{case-study:consolidation}. We utilize the motivation presented in Blink~\cite{wang2020blink} which highlighted that there is bandwidth imbalance between GPUs within a node, \eg bandwidth between GPU 0 and GPU 3 is twice that of bandwidth between GPU 0 and GPU 1 for \emph{p3.8xlarge} machines.
To improve bandwidth utilization we introduce a bandwidth aware intra-node placement policy which maximizes the aggregate bandwidth for multi-GPU jobs, \ie place multi-GPU jobs on GPUs on high bandwidth pair.
To support this bandwidth aware intra-node placement policy, in \system we only needed to add 14 additional lines of code to implement this policy.
To evaluate our Intra-Node Placement Policy we used the \phtrace{} and tracked the avg bandwidth experienced by single node, multi-GPU jobs. For the experiment, we used FIFO scheduler with consolidation as global placement policy. As shown in Table~\ref{tab:avg_bw_intranode} our Intra-Node Placement policy improves bandwidth observed by $1.47\times$.

\begin{table}[t]
\caption{\small{\textbf{Evaluating Bandwidth Aware Intra-Node Policy:} The new policy improves observed bandwidth by around $1.4\times$}}
\vspace{-10pt}
\label{tab:avg_bw_intranode}
\resizebox{\linewidth}{!}{
\begin{tabular}{@{}ll@{}}
\toprule
Policy                                         & Avg Bandwidth Observed (Gbps) \\ \midrule
\multicolumn{1}{l|}{Random}                    & 58.7                          \\
\multicolumn{1}{l|}{Bandwidth Aware Placement} & 86.5                          \\ \bottomrule
\end{tabular}}
\vspace{-5pt}
\end{table}

\noindent\textit{Takeaway: \system{} is flexible and can support adding new policies with a few lines to code enabling rapid prototyping of new schedulers.}

\section{Blox Implementation}
\label{sec:design}

\begin{table*}[t]
\caption{\small{A list of key abstractions and their possible implementations for composing DL schedulers in \system{}.}}
\vspace{-10pt}
\label{tab:possible_instance}
\resizebox{\linewidth}{!}{
\begin{tabular}{lc}
\hline
\textbf{Abstraction}                                      & \textbf{Possible Instances}                                                                                                         \\ \hline
\multicolumn{1}{l|}{Job Admission Policy}       & \multicolumn{1}{c}{user job quota, user resource quota, job type quota, job resource quota}                               \\ \hline
\multicolumn{1}{l|}{Cluster Management}         & \multicolumn{1}{c}{add/remove nodes, maintain machine map (job-resource mapping, and resource free list)}                                                                   \\ \hline
\multicolumn{1}{l|}{Job Scheduling Policy}      & \multicolumn{1}{c}{FIFO, FIFO + Priority, LAS, SRTF, maximize throughput, discreet LAS, largest marginal gain,  FTF (Themis), heterogeneity-aware (Gavel), Pollux}                                  \\ \hline
\multicolumn{1}{l|}{Job Placement Policy}       & \multicolumn{1}{c}{first available, maximize consolidation, application determined placement, min network interface}      \\ \hline
\multicolumn{1}{l|}{Job Launch Mechanism}       & \multicolumn{1}{c}{zipfile, command line, docker}                                                                         \\ \hline
\multicolumn{1}{l|}{Job Preemption and restart} & \multicolumn{1}{c}{CRIU, iteration boundary, run to completion,}                                                          \\ \hline
\multicolumn{1}{l|}{Metric Collection}          & \multicolumn{1}{c}{per-iteration time, loss, finish time fairness estimate, throughput, inference requests per unit time} \\ \hline
\end{tabular}
}
\vspace{-10pt}
\end{table*}

In previous sections we gave examples of using \system to build and evaluate existing
schedulers on a common footing and to support building new schedulers and policies.

We will open source \system and all the implemented schedulers for the benefit the community.
In this section, we present an overview of \system, describe its key design philosophy and our implementation.

\subsection{\system Design Overview}
\system is designed with the insight that DL schedulers can be composed by using different instances of a subset of abstractions. As long as the inputs and outputs of these abstractions are maintained, the users can create news instances of these abstractions. Table~\ref{tab:possible_instance} lists few different instances which are possible of the abstractions present in \system. Further, users can create their own additional abstractions and chain them with other abstractions in a similar way.

We also believe that to create an instance of any abstraction or a new abstraction, the user only needs access to the cluster state - which includes node types, gpu types, memory utilization, disk utilization and compute utilization and state of jobs - which includes job type, resource requirements, run time metrics like per iteration time, gpu memory needed, disk space needed to name a few. With this information users can create both new instances of existing abstractions and new abstractions as well. To provide access to this information \system provides two well defined data structures \texttt{JobSate} and \texttt{ClusterState}, \texttt{JobState} provides access to both completed jobs and currently active jobs to the user and all metrics associated with jobs resource requirement and run time information. We provide more details of these data structures in Appendix~\ref{appendix:api_data_structure}.

\subsection{\system API Design}
\label{sec:api}
\system is designed to provide flexibility to the user. In general each abstraction in \system takes atleast two inputs,
the two information data structures- \texttt{JobState} and \texttt{ClusterState}- beyond these two inputs each of these abstractions take additional inputs, \eg as shown in Figure~\ref{fig:blox_chain_flow} \textit{job admission policy} takes the new jobs arrived as well as the \texttt{JobState} and \texttt{ClusterState} and outputs jobs that should be accepted to schedule on the cluster.
Further the abstractions have a well defined output which is usually fed into next set of abstractions. We provide API details for each of the abstractions present in \system in Appendix~\ref{appendix:api_data_structure}.

\subsection{Implementation}
\begin{figure}[t]
    \includegraphics[width=\linewidth]{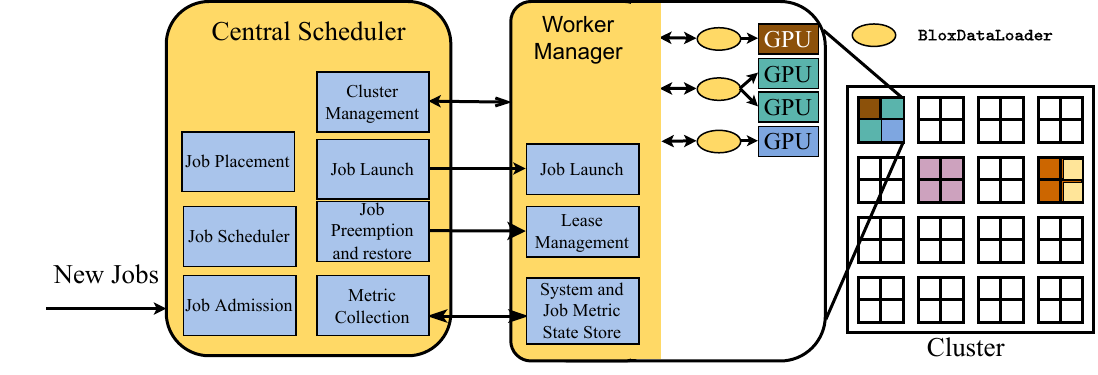}
    \centering
    \vspace{-20pt}
    \caption{\small{\textbf{\system{} Implementation:} consists of three major components: a \cscheduler{}, \wmanager{} on each worker and a \blcl{} that links to DL jobs. Arrows show RPC communication used by \system{} for initialization, job launch, preemption, and metric collection.}}

    \label{fig:blox_design}
    \vspace{-15pt}
    \end{figure}

Having described the key abstractions necessary for building DL schedulers and exploring some case studies, we next describe additional details of our implementation. \system is implemented in around 8000 lines of Python and we use gRPC for communication between our distributed system components~\cite{grpc}.
Similar to prior centralized scheduling frameworks~\cite{vavilapalli2013yarn, hindman2011mesos, verma2015large}, we build \system{} in three high level modules (Figure~\ref{fig:blox_design}). \cscheduler, where much of the scheduling logic runs, \wmanager that runs on each node and manages the node, and \blcl used by DL training jobs to interact with \system.

\paragraph{\cscheduler.} Similar to existing DL schedulers, we use a centralized process to perform scheduling and resource management decisions. \cscheduler encapsulates all the functionalities needed for centralized decision making and instantiates all the modules related to job scheduling, placement decisions and cluster management.

\label{sec:sys_components}

\begin{table*}[t]
\caption{\small{\textbf{Input and Outputs to abstractions:} The Table lists the input and outputs to each of the abstraction present in a DL scheduler.}}
\label{tab:api_design}
\resizebox{0.8\linewidth}{!}{%

\begin{tabular}{@{}lll@{}}
\toprule
Abstraction                                & Input                                                          & Output                                                                                                                           \\ \midrule
\multicolumn{1}{l|}{Job Admission Policy}  & \multicolumn{1}{l|}{new-jobs, ClusterState, JobState}          & accepted-jobs                                                                                                                    \\ \midrule
\multicolumn{1}{l|}{Cluster Management}    & \multicolumn{1}{l|}{new-nodes, ClusterState}                   &                                                                                                                                  \\ \midrule
\multicolumn{1}{l|}{Job Scheduling Policy} & \multicolumn{1}{l|}{ClusterState, JobState}                    & \begin{tabular}[c]{@{}l@{}}job-priority-list \\ (sorted by priority to schedule)\end{tabular}                                    \\ \midrule
\multicolumn{1}{l|}{Job Placement Policy}  & \multicolumn{1}{l|}{job-priority-list, ClusterState, JobState} & \begin{tabular}[c]{@{}l@{}}job-allocations (job ids and gpus to launch on),\\  job-preemptions (job ids to preempt)\end{tabular} \\ \midrule
\multicolumn{1}{l|}{Job Launch Mechanism}  & \multicolumn{1}{l|}{job-allocations,  ClusterState, JobState}  &                                                                                                                                  \\ \midrule
\multicolumn{1}{l|}{Job Preemption}        & \multicolumn{1}{l|}{job-preemptions, ClusterState, JobState}   &                                                                                                                                  \\ \midrule
\multicolumn{1}{l|}{Metric Collections}    & \multicolumn{1}{l|}{JobState, ClusterState}                    &
\end{tabular}%
}
\vspace{-10pt}
\end{table*}

\paragraph{Implementation changes} In Table~\ref{tab:changes_made} we provide an overview of abstractions updated to implement each scheduler. We show we are able to use reuse large parts of code for building a scheduler.

\paragraph{\wmanager.} A \wmanager runs on every server in the cluster to manage operations on the machine and
execute the decisions made by the \cscheduler (e.g., job launch, preemption, etc.). \wmanager also acts as local state store for applications to push metrics which will be used by scheduler in future decision making. \wmanager's also obtain a lease from the \cscheduler when a new job is assigned to a worker. We discuss how lease renewal and revocation works in detail below.

\paragraph{\blcl.} As DL schedulers use application-specific metrics for scheduling, we need a client library that applications can use to collect these metrics. Furthermore, supporting iteration-level preemption of DL training also requires integration between the applications and \system{}. We design \blcl to address these two requirements. \blcl is composed of two components \blit and \texttt{WorkerMetricsCollector}.

\blit is as a wrapper over the native PyTorch or Tensorflow dataloader, and

it enables our lease based preemption mechanism. \blit checks the lease status with the \wmanager at each iteration and if the lease is not available the application is preempted by taking a consistent checkpoint.

\texttt{WorkerMetricsCollector} allows applications to provide the \cscheduler, via the \wmanager{}'s metrics state store, with relevant job-related metrics at runtime.  The \texttt{WorkerMetricsCollector} interface accepts a generic key-value pair from applications and thus allows them to push any arbitrary application metric like loss, norm of gradients, validation accuracy, etc., that can be used by the \cscheduler{}.

An important aspect of our implementation is designing data structures which can supply \system{} abstractions with information about all the jobs and the cluster.
 Our goal was to design data-structures
 that are flexible enough to track all the information but still support fast queries. To that end we chose to store the \texttt{ClusterState} in a data frame which allows easy filtering and querying regarding the status of machines. To store job related information we designed \texttt{JobState} where all the information is kept in a dictionary-like data structure, and provides users with the flexibility of tracking any information related to a job.
\subsection{\system Dataflow and API}
\label{appendix:api_data_structure}

\paragraph{Data Structures:}
In \system we maintain two core data structures, \texttt{ClusterState}, \texttt{JobState}. These are implemented as \textit{python} classes. \texttt{ClusterState} provides access two state variables, one is a dictionary which keeps information about each node type in the cluster with information like CPU type, Memory, Network bandwidth, interconnect bandwidth. The second is a tabular data structure, which which has a row for each GPU on the cluster. The columns in this tabular data structure are (i) node-id (which represents the id of the node which the GPU is on), (ii) global gpu-id (an increasing counter which ID of each GPU), (iv) local GPU-ID (represents the gpu id with respect to current node) (iii) gpu-type (the type of GPU) (iv) state of GPU (running, free) (v) free-memory (memory free on the GPU) (vi) jobs running (list of jobs running on the GPU).

The second data structure is \texttt{JobState}. It provides access to a state variable, which keeps track of each job which is submitted but has not finished.
All the information about the job including type, launch command, preferences, metrics associated with a jobs, iteration time to name a few are tracked in this data structures. Another state variable, keep track of metrics of jobs which have finished like completion time, resources used etc.

These two data structures provides complete state of the jobs and the cluster. We believe with access to these two datastructures a user can write a new policy.

\begin{table*}[t]
\caption{\small{\textbf{Details of abstractions and changes made:} We provide details of changes made in each abstraction to implement a scheduler.}}
\vspace{-5pt}
\label{tab:changes_made}
\resizebox{\linewidth}{!}{
\begin{tabular}{@{}lll@{}}
\toprule
Scheduler                 & Abstractions Modified & Changes Made                                                                                                                                                                                                              \\ \midrule
LAS                       & Scheduling Policy     & - Sorted Jobs by service attained                                                                                                                                                                                         \\ \midrule
\multirow{2}{*}{Tiresias~\cite{gu2019tiresias}} & Scheduling Policy     & \begin{tabular}[c]{@{}l@{}}- Add configurable number of queues and discreet LAS\\ - FIFO within queues and LAS across queues\end{tabular}                                                                                 \\
                          & Placement Policy      & - Assign jobs based on their placement preference, choosing between consolidated vs unconsolidated placement preference                                                                                                   \\ \midrule
\multirow{3}{*}{Optimus~\cite{peng2018optimus}}  & Scheduling Policy     & \begin{tabular}[c]{@{}l@{}}- Assign one GPU to each job in expected convergence order\\ - If GPUs still free then assign additional GPUs based on expected convergence speedups\end{tabular}                              \\
                          & Placement Policy      & - Prefer consolidated placement                                                                                                                                                                                           \\
                          & Metric Collection     & - Add additional key to collect loss value per iteration                                                                                                                                                                  \\ \midrule
\multirow{3}{*}{Gavel~\cite{narayanan2020heterogeneity}}    & Scheduling Policy     & - Implemented Gavels Optimization based routine which outputs share for  each GPU types for LAS                                                                                                                           \\
                          & Placement Policy      & - Implemented Gavels Placement Algorithm ()                                                                                                                                                                               \\
                          & Metric Collection     & - Push additional key to update the iteration time observed                                                                                                                                                               \\ \midrule
\multirow{3}{*}{Pollux~\cite{qiao2021pollux}}   & Scheduling Policy     & \begin{tabular}[c]{@{}l@{}}- Implement the Goodput optimizing scheduling and placement policy \\ - Pollux makes both scheduling and placement decisions together, we combine scheduling and placement policy\end{tabular} \\
                          & Workload Generation   & - Pollux uses a custom workload generation, and also requires additional parsers to read profiled data about jobs                                                                                                         \\
                          & Metric Collection     & - Update Metric Collection to collect running goodput at each iteration                                                                                                                                                   \\ \midrule
\multirow{2}{*}{Themis~\cite{mahajan2020themis}}   & Scheduling            & - Implement finish time fairness scheduler                                                                                                                                                                                \\
                          & Metric Collection     & - Collect fair share during each round duration for the scheduler to use during next round                                                                                                                                \\ \midrule
\multirow{2}{*}{Synergy~\cite{mohan2021synergy}}  & Scheduling            & - Modify the scheduler to use Synergy scheduling policy both Proportional and Synergy-Tune                                                                                                                                \\
                          & Placement Policy      & - Modify placement policy to account for CPU and Memory resources while performing placement                                                                                                                              \\ \bottomrule
\end{tabular}}
\vspace{-10pt}
\end{table*}

\paragraph{API for Abstractions:}
In Table~\ref{tab:api_design} we provide details of inputs and outputs to each abstractions.
For each of these abstractions we provide a base class template which can also except additional arguments beyond just the ones listed as key word arguments. The users can modify or create new instance of any existing abstraction.

Each abstraction in \system is designed to access our two data structures which can provide all the information that a user can use in order to write a new abstraction or a new instance of existing abstraction.

\section{Evaluation}
\label{sec:eval}

\begin{figure}[t]
    \centering
    \begin{minipage}[t]{0.22\textwidth}

    \includegraphics[width=0.99\linewidth]{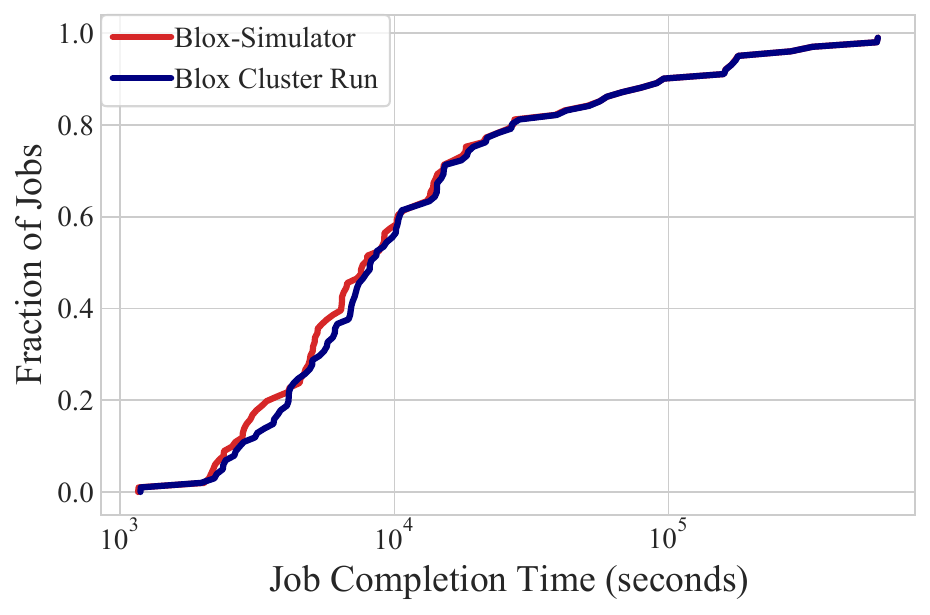}
    \vspace{-15pt}
    \caption{\small{\textbf{\system{} simulator fidelity:} Average JCT from simulator compared against an actual run on cluster using the \sysname{} runtime.}}
    \label{fig:simulator_fidelity}
    \vspace{-15pt}
    \end{minipage}\quad
    \begin{minipage}[t]{0.22\textwidth}
    \includegraphics[width=0.99\linewidth]{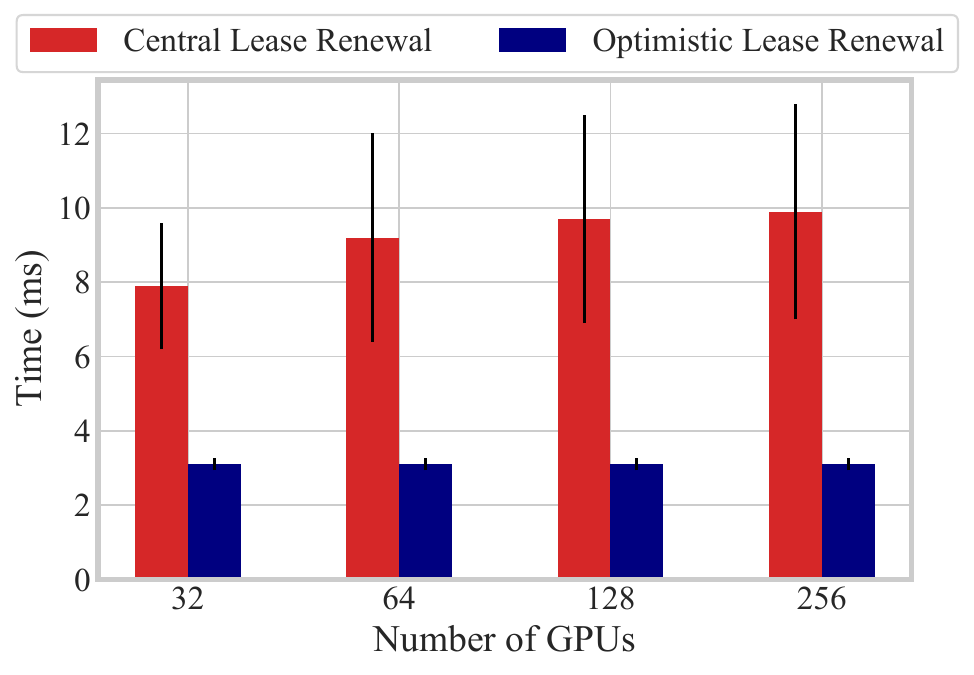}
    \vspace{-15pt}
    \caption{\small{\textbf{Optimisitc vs. Centralized Lease Renewals}. Optimistic lease renewals are faster and more scalable.}}
    \label{fig:opt_lease_renew}
    \vspace{-15pt}
    \end{minipage}

\end{figure}

\paragraph{Blox Implementation Fidelity.} We next compare Blox's simulator with its deployment runtime implementation on a 32 GPU ($8\times$ \emph{p3.8xlarge}) Amazon EC2 cluster.  we compare \system's simulator and actual cluster runs by plotting the CDF of job completion times on a trace of 100 jobs arriving at the load average of 4 jobs per hour. We use the FIFO scheduling policy and First-Free GPU placement policy.
We ensure that the simulator can capture the job launch and preemption overheads and profile these overheads for the models we use (Table~\ref{tab:models}). From Figure~\ref{fig:simulator_fidelity} we see that the CDFs are very similar with the 25th, 50th and 75th values of the two distributions differing by 1.7\%, 5.8\% and 2.2\% respectively.
From a per-job perspective, we found that the average difference in JCT is around 6.1\%

This shows that \system{} can be used by researchers to develop new schedulers using simulations and then transparently validate that on real-world clusters.

\paragraph{Leases for Preemption}
It is common for round-based DL schedulers to use centralized lease-based mechanisms to aid in job preemption~\cite{narayanan2020heterogeneity,mohan2021synergy}. We first discuss centralized lease checks and then provide details about how we improve upon it using our optimistic lease renewal policy.
With centralized lease checking, workers for each job typically need to check with a centralized entity if their lease can be extended for another round or if they  are to be preempted at the end of the current round.  However, centralized lease checking scales poorly with the number of accelerators and jobs in the cluster (e.g., as shown Figure~\ref{fig:opt_lease_renew}).

To address the overheads with centralized lease checks we propose using \textit{optimistic lease renewals} in \system.  Here, we assume leases are automatically renewed unless the the \cscheduler revokes the lease with the \wmanager (when it wants to preempt a job). Once an iteration completes, the \blit within each job will check its lease status with the local \wmanager, thereby eliminating the need for periodic lease checks to the \cscheduler. % traffic.

When preempting distributed jobs, there can be a deadlock due to lease revocation reaching different workers at different times. This could lead to some workers proceeding with the next iteration while other workers deciding to terminate, causing deadlocks and inconsistent checkpoints.
To solve this problem we use a two phase lease expiration mechanism, allowing the distributed workers to coordinate among themselves and reach a consensus on when it is safe to terminate.
The \cscheduler sends the lease revocation signal to only one of the workers (say worker $w$). $w$ checks the current iteration number ($i$) and marks the job to be preempted after the next iteration ($i+1$). Next, $w$ synchronously propagates the exit iteration number to all other workers before it begins iteration $i+1$; \footnote{In the worst case, even if all other workers would have raced ahead to the end of iteration $i+1$, they would wait for $w$ at a collective call (e.g., AllReduce) at the end of the iteration}. Following this, all the workers exit in tandem at the end of iteration $i+1$. This leads to consistent checkpoints and avoids deadlock. The only drawback of this approach is that the job exit is delayed by one iteration. However, since the iteration time is significantly smaller than the round duration, this delay is inconsequential in practice.

\paragraph{Evaluating lease renewal overheads.}
To evaluate the benefits of \emph{optimistic lease renewals} we also modified \system to implement \emph{central lease renewal}, \ie each job checks the lease status with the centralized scheduler.

To compare their performance scalability, we vary the number of GPUs available in the cluster.
In Figure~\ref{fig:opt_lease_renew} we see that \emph{optimistic lease renewal} is more than 50\% faster than \emph{central lease renewal}.
Further we also observe that the time taken for \emph{optimistic lease renewal} remains constant while the time for \emph{central lease renewal} grows as we scale the number of GPUs, highlighting the performance bottleneck of the centralized scheme.

\section{Discussion}
\label{sec:discussion}

\paragraph{Experience using \system}
To evaluate the usability of \system, we also invited two group of graduate students to re-implement existing schedulers using \system{}. One group re-implemented  Themis~\cite{mahajan2020themis} while another group re-implemented Optimus \cite{peng2018optimus}; these were independent from our implementations of Themis and Optimus. These groups did not have any prior experience in building DL schedulers and started with the FIFO scheduler in \system{}. Once the students had read the corresponding prior research papers, each group reported that they were able to re-implement these schedulers in \system in around \emph{40 hours} (4-5 days) of work. We also made improvements to \system{} based on their experience, including additional documentation, better error handling, and improved support for parsing new workload traces. The aspect of \system which the students liked the most was that once they figured out scheduling and placement logic, the framework helped them run simulations and experiments very quickly.  Encouraged by this experience, we intend to continue using \system{} for such student projects and course assignments.

\paragraph{Limitations of Simulation in Scheduling Research}

It is common practice in scheduler research~\cite{qiao2021pollux,boutin2014apollo,crankshaw2017clipper,narayanan2020heterogeneity,mohan2021synergy} to validate the fidelity of the simulator by comparing the results obtained in simulation with real-world runs for a specific workload trace (typically at smaller scales), and then using simulations to sweep various parameters for scheduler evaluation (including larger-scale runs).
Simulations provide an effective way to evaluate innovations at larger scales without requiring access to expensive large-scale deployments.
 Simulation are natively supported in \system. To minimize variance between real cluster runs and simulations, \system uses the  same code path for simulations as for real cluster runs.
 However, simulations can have some differences from real cluster results, due to variability in hardware~\cite{sinha2022not}, overlooking additional system aspects like disk loading times and resource contention. Notwithstanding these limitations, we believe simulations provide indispensable insights, allowing users to balance the trade-off between accuracy of the experiment and cost-effectiveness through detailed evaluations.
\paragraph{Beyond ML Training}
While our discussion so far has been focused on schedulers for DL training jobs, in the future we plan to investigate if \system can also be used to support inference schedulers and hyper-parameter tuning libraries.
To study the potential for supporting inference schedulers, we consider Nexus~\cite{shen2019nexus}, a recent work that improves efficiency of inference while supporting multiple models and applications. We detail our implementation of Nexus in \system in Appendix~\ref{appendix:nexus}.

For hyper-parameter tuning we consider algorithms such as HyperBand~\cite{li2017hyperband} as a scheduling algorithm, where the hyper-parameter optimization algorithm chooses which subset of configurations should continue running based on training progress. We can implement HyperBand's job pruning logic as a scheduling policy and modify \blcl to propagate training progress to \cscheduler{}.
\paragraph{Joint Scheduling, Placement and Admission control}
One potential limitation of decomposing schedulers into different components is that each module has to make decisions without control over the other modules.
In the context of ML training schedulers, some policies like AntMan~\cite{xiao2020antman} have a scheduling policy that first evaluates if placement constraints can be met for a job before allocating it resources. Similarly, in inference schedulers like Nexus, the scheduling policy of how many GPUs should be allocated to each model also acts as an admission control policy to determine which models can be supported without missing SLOs. Having admission control be done before scheduling could lead to sub-optimal scenarios where not all GPUs are efficiently used.
Such scenario can be handled by defining a combined module that performs both operations (e.g., scheduling and placement) and inserting the module in the appropriate part of the workflow. The flexible composition logic in \system{} where state is passed through shared data structures in \texttt{ClusterState} and \texttt{JobState} allows developers to define a different scope for new modules while integrating with existing modules. % that are already present.

\paragraph{Round-based vs. Churn-based Scheduling} \system{} currently supports schedulers which follow a centralized round-based mechanism for scheduling. While round based scheduling is the most common design used by DL training schedulers, prior research in datacenter scheduling have also proposed decentralized designs~\cite{ousterhout2013sparrow, boutin2014apollo} and schedulers that perform allocation only when new jobs arrive~\cite{yoo2003slurm} or when configuration changes (i.e., churn-based scheduling). While our optimistic lease renewal can be used to support scheduling policies where the scheduling loop only kicks in on churn, we leave such investigation to the future.

\paragraph{Support for hybrid and distributed Schedulers.}

\system can also potentially support distributed and hybrid Schedulers. For performing distributed scheduling like in Omega \cite{schwarzkopf2013omega}, there could be multiple ``centralized schedulers'' running in parallel, each having a copy of the \texttt{ClusterState}.
One would need to modify the node manager to handle conflicts and choose the appropriate job to run in case of conflicts. Our get\_metrics call can update all copies of the \texttt{ClusterState} providing all schedulers with an accurate state of the cluster periodically.
\system{} can also support hybrid architectures similar to Apollo~\cite{boutin2014apollo}. In this case we could have a single centralized scheduler with multiple Job Scheduling abstractions running in parallel (e.g., Python multi-process), sharing a global view of the \texttt{ClusterState}.

\section{Conclusion}
\label{sec:conclusion}
We presented \system, a modular toolkit to allow researchers and practitioners compare, compose and build new DL schedulers. \system provides a set of extensible building blocks which can be easily modified to implement new and existing schedulers. We showcased the generality of \system by implementing \emph{7} existing schedulers and validated our implementations by reproducing results from prior work. We also performed a number of case studies to highlight how \system{} can be used to better understand existing schedulers under new scenarios (cluster load, hardware, models), and how we can quickly prototype new designs by composing or creating new modules. We hope that \system will be a resource that the systems research community can use to rapidly build and evaluate research DL schedulers in the future.

\clearpage
\bibliographystyle{plain}
\bibliography{ref}
\clearpage
\appendix
\section{Automatic Scheduler Synthesizer}
\begin{figure}[t]
    \centering
    \includegraphics[width=0.9\linewidth]{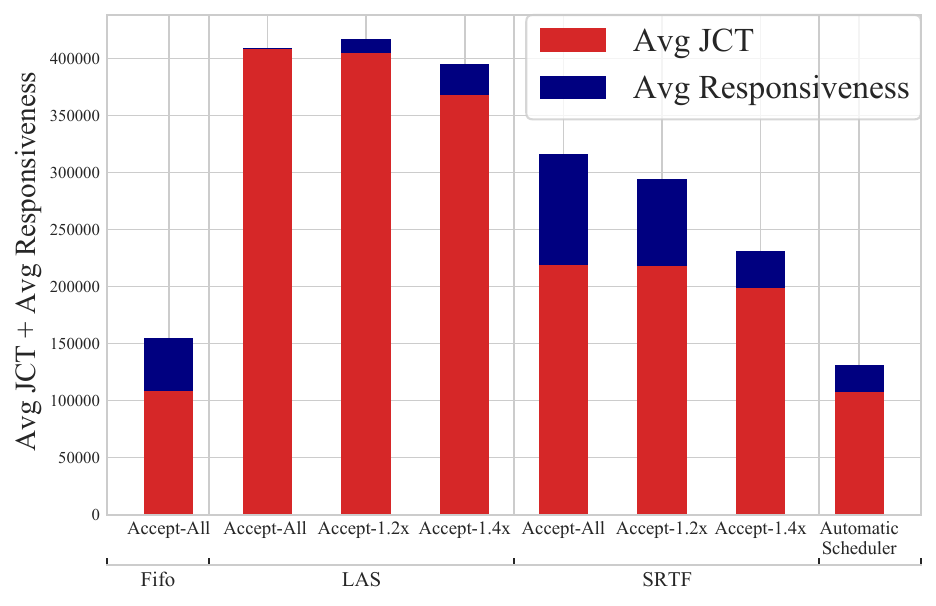}
    \vspace{-10pt}
    \caption{\small{\textbf{Comparing Responsiveness and JCT:} Automatic synthesizer is able to minimize both average JCT and responsiveness.}}
    \label{fig:dynamic_policy_responsive}
    \vspace{-10pt}
\end{figure}

\begin{figure}[t]
    \centering
    \includegraphics[width=0.9\linewidth]{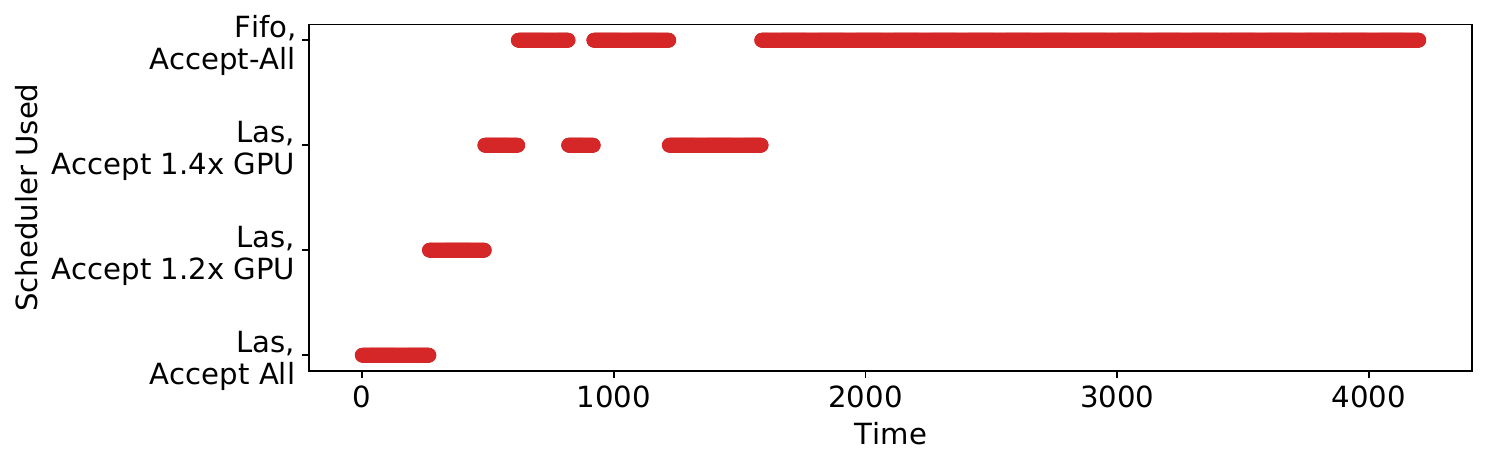}
    \vspace{-10pt}
    \caption{\small{\textbf{Scheduler used by automatic synthesizer:} The temporal distribution of policies used by automatic synthesizer shows that initially it first used LAS, than LAS with $1.2\times$ acceptance policy and then eventuall $1.4\times$ acceptance policy and finally transitioning to FIFO. LAS was chosen because it was reducing responsiveness while not hurting JCT, but eventually as jobs keep getting preempted JCT started increasing and our automatic scheduler switched to FIFO. }}
    \label{fig:temporal_policy_responsive}
    \vspace{-10pt}
\end{figure}

In this section we present additional results for automatic synthesizer. We show that our setup is capable of minimizing multiple objectives. In Figure~\ref{fig:dynamic_policy_responsive} we show that our Automatic Synthesizer is able to minimize both Avg JCT and Avg Responsiveness.
To perform this we use the same technique of running simulations in parallel and then calculating both responsiveness and jct one every ten rounds. We choose the option which minimizes both these values simultaneously.
\label{appendix:additional_automatic}
In Figure~\ref{fig:temporal_policy_responsive} we show the temporal distribution of policies chosen by the Automatic Scheduler Synthesizer.

\section{\system API and Programming Model}
With  the aid of Figure~\ref{fig:blox_chain_flow} we describe the programming model of \system.

\paragraph{\system Data Structures.}

To track information related to Jobs and the Cluster, \system uses two data structures - \texttt{JobState} and \texttt{ClusterState}. \texttt{JobState} is a flexible key-value data structure, \ie, dictionary style structure which keeps track of all job related information and any other metadata the user wants to track for a job. For \texttt{JobState} flexibility is the most important characteristic, different schedulers track various different metrics; for example  Synergy~\cite{mohan2021synergy} tracks different GPU, CPU configurations and slowdowns associated with them, Gavel~\cite{narayanan2020heterogeneity} tracks throughput for different packing scenarios for hardware types present on the cluster and Optimus~\cite{peng2018optimus} tracks loss value of  applications. Therefore, we choose a key-value style data-structure to represent \texttt{JobState.} For \texttt{ClusterState} we choose a column data structure, \ie, data-frames or SQL tables. The fields to track for \texttt{ClusterState} are limited and query speed is significantly more important. In future, we also aim to provide additional flexibility in the \texttt{ClusterState} data-structure by allowing users to track arbitrary metrics using Binary Large Object support present in modern SQL systems.

These two data structures provide a user with a full view of the state of the jobs and state of the cluster in the scheduler. We believe using the information present in these two data-structures, users can implement any scheduling logic. In Figure~\ref{fig:blox_chain_flow} at Line 8 and 9 we initialize the cluster state and job state.

\paragraph{\system Modules.}

We first describe \system modules. As stated in Section~\ref{sec:blox_overview} and Figure~\ref{fig:sched_schem}, DL schedulers are built using discrete modules. \system provides abstractions to implement these modules, \eg, Admission Policy, Scheduling Policy, Placement Policy (discussed in Section~\ref{sec:api} and~\ref{sec:blox_overview}. The core of idea of \system is that we can easily reuse the code across different schedulers by making only changes to the specific abstractions. In Figure~\ref{fig:blox_chain_flow} in Line 4,5 and 6 we initialize the abstractions. In Line 24, 30 and 33 we execute the realization of these abstractions to run the scheduler.

\paragraph{Chaining Modules in \system.}
As discussed in Section~\ref{sec:blox_overview}, DL schedulers can be implemented as a chain of abstractions. Figure~\ref{fig:blox_chain_flow} shows an example of that. We keep running the scheduler until the specified exit condition is satisfied. To build a new scheduler, the user has to modify the relevant abstraction and replace it while initializing as in Line 4,5 or 6. Based on new additions the user can modify the scheduling loop and chain the abstractions based on their design.

\paragraph{Adding New Modules to \system.}
We understand that a user in future will likely want to add new abstractions. \system is well equipped for such a use case. To provide information about the jobs and cluster state, users can use existing data-structures for \texttt{JobState} and \texttt{ClusterState}. When chaining the new abstraction, users will have to determine where the new abstraction will be in the scheduling flow and then insert it there. We note that the changes only have to be abide by the inputs and outputs of the abstractions between which the new abstraction is being inserted. For example, say a user creates a new abstraction called \emph{ShinyNewPolicy} which takes in output of the scheduling policy and makes changes to it, before sending it to the placement policy. When writing \emph{ShinyNewPolicy} the user needs to be aware of the output of the scheduling policy and the format in which it is delivered, and correspondingly the input format of the placement policy.

\section{Additional Discussion}
\label{appendix:nexus}
\paragraph{Implementing Nexus in \system}
Nexus is composed of three components: frontends, backends and global schedulers. The frontends are responsible for receiving inference requests and routing it to the appropriate backend for inference. Backends are GPU servers which host the model for inference and process received requests. The global scheduler acts as the control plane; it instructs backends on which models should be loaded and the batch size to use for each of them. The global scheduler also provides frontends with routing tables that indicate which backend a request should be routed to.
We can implement Nexus's global scheduler in our scheduling policy abstraction. The input to our scheduling policy would be the number of requests received at the frontends and this can be shared using the \blcl. The scheduling policy can implements Nexus' \textsc{SquishyBinPacking} algorithm to compute the number of GPUs and the batch size for each GPU while ensuring that inference requests can meet their SLOs. After the scheduling policy completes, we can use the lease extension mechanism to install the new routing table at the frontends.
We ere able to design a prototype implementation currently using \system{}, however our current architecture does not support propagating batch size configuration changes at a fine granularity. To support such applications in the future, we plan to study if we can generalize the communication between the \cscheduler{} and \wmanager{} so as to  rapidly change configurations, routing logic etc.

\end{document}